\documentclass[aps,prd,superscriptaddress,onecolumn,floatfix,nofootinbib]{revtex4}
\usepackage{amsmath,graphicx,hyperref}

\pdfoutput = 1
\graphicspath{{figs/}}

\newcommand\one{\leavevmode\hbox{\small1\normalsize\kern-.33em1}}

\newcommand{\qqquad}{\qquad \qquad}

\newcommand{\gev}{\text{GeV}}
\newcommand{\tev}{\text{TeV}}

\def\slashchar#1{\setbox0=\hbox{$#1$}           % set a box for #1
   \dimen0=\wd0                                 % and get its size
   \setbox1=\hbox{/} \dimen1=\wd1               % get size of /
   \ifdim\dimen0>\dimen1                        % #1 is bigger
      \rlap{\hbox to \dimen0{\hfil/\hfil}}      % so center / in box
      #1                                        % and print #1
   \else                                        % / is bigger
      \rlap{\hbox to \dimen1{\hfil$#1$\hfil}}   % so center #1
      /                                         % and print /
   \fi}

\newcommand{\eg}{\textsl{e.g.}\;}

\newcommand{\be}{\begin{eqnarray*}}
\newcommand{\ee}{\end{eqnarray*}}

\newcommand{\bee}{\begin{eqnarray}}
\newcommand{\eee}{\end{eqnarray}}
\newcommand{\beeq}{\begin{equation}}
\newcommand{\eeeq}{\end{equation}}

\newcommand{\eps}{\varepsilon}
\begin{document}

\title{Resonance Searches with an Updated Top Tagger}

\author{Gregor Kasieczka}
\affiliation{Institute for Particle Physics,
             ETH Z\"urich, Switzerland}

\author{Tilman Plehn}
\affiliation{Institut f\"ur Theoretische Physik, 
             Universit\"at Heidelberg, Germany}

\author{Torben Schell}
\affiliation{Institut f\"ur Theoretische Physik, 
             Universit\"at Heidelberg, Germany}

\author{Thomas Strebler}
\affiliation{Institute for Particle Physics,
             ETH Z\"urich, Switzerland}

\author{Gavin P. Salam}
\affiliation{CERN, PH-TH, CH-1211 Geneva 23, Switzerland}

\date{\today}

\begin{abstract}
  The performance of top taggers, for example in resonance searches,
  can be significantly enhanced through an increased set of variables,
  with a special focus on final-state radiation. We study the
  production and the decay of a heavy gauge boson in the upcoming LHC
  run. For constant signal efficiency, the multivariate analysis
  achieves an increased background rejection by up to a factor 30
  compared to our previous tagger. Based on this study and the
  documentation in the Appendix we release a new
  HEPTopTagger2 for the upcoming LHC run. It now includes an
  optimal choice of the size of the fat jet, N-subjettiness, and
  different modes of Qjets.
\end{abstract}

\maketitle
\tableofcontents
\newpage

%%%%%%%%%%%%%%%%%%%%%%%%%%%%%%%%%%%%%%%%%%%%%%%%%%%%%%%%%%%%%%%%%%%%%%
\section{Introduction}
\label{sec:intro}

After the discovery of the Higgs boson, a keystone of the Standard
Model, one main task for the upcoming LHC runs will be searches for
physics beyond the Standard Model. Several open experimental and
theoretical questions point to additional particles or structures at
energies above the electroweak energy scale~\cite{review}.  A very
generic feature of many extensions of the Standard Model is the
presence of additional heavy particles which preferentially decay to a
pair of top quarks~\cite{tt_resonances}. One example for such a
resonance could be a heavy neutral $Z'$-gauge boson with a TeV-scale
mass. Historically, such states were only searched for using
semi-leptonically decaying top pairs. There, a kinematic
reconstruction is based on an approximate reconstruction of the
missing neutrino momentum through a $W$-mass or top mass
condition. In the last LHC run this search channel was supplemented by
resonance searches based on boosted, hadronically decaying top pairs.
In the corresponding ATLAS analysis~\cite{atlas} the
\textsc{HEPTopTagger}~\cite{heptop1,heptop2} and the template
tagger~\cite{template} each showed a similar reach, comparable with
the semileptonic channel. This experimental success is based on rapid
progress in the field of just substructure both experimentally and
theoretically, which will gain even more momentum during the 13~TeV
LHC run.\bigskip

The field of top and Higgs tagging~\cite{tagging_review} started
essentially as a Gedankenexperiment to illustrate recombination jet
algorithms~\cite{seymour}. After some early attempts for example to
tag hadronically decaying tops~\cite{early} it took off with the
development of the BDRS Higgs tagger with its mass drop
condition~\cite{understanding} and a filtering step targeting
underlying event and pile-up~\cite{bdrs}.  The first top taggers were
simple, deterministic algorithms which could identify and reconstruct
hadronically decaying top quarks including subjet
$b$-tagging~\cite{hopkins,heptop1,soft_drop,mass_jump}. They were
based on deliberately simple structures and algorithms, to firmly
establish subjet methods in ATLAS and CMS.
After the experimental success of these completely new analysis tools
in the first run of the LHC, the upcoming run will benefit from more
advanced top tagging methods. Those include multivariate
taggers~\cite{heptop4}, template taggers~\cite{template}, as well as
shower deconstruction~\cite{shower_deco} or event
deconstruction~\cite{event_deco}\footnote{Why a kinematic selection as
  naive as `top buckets'~\cite{buckets} also seems to work is beyond
  the comprehension of the authors.}. For those specialized tools the
challenge will be to still provide a universal top tagging approach,
which on the one hand allows for optimal experimental results, but on
the other hand identifies and reconstructs boosted top quarks
independent of the specialized analysis framework.\bigskip

Over time, the original \textsc{HEPTopTagger}~\cite{heptop1} has gone
through several rounds of improvements. The first modification
included a re-formulation of the algorithm, leading to the trademark
A-shaped kinematic cuts~\cite{heptop2}. One of the key observations
leading to these cuts is that in the absence of a $b$-tag it is not
helpful to uniquely identify the two $W$-decay jets because in typical
top decays there will be two jet-jet combinations which reconstruct to
an invariant mass around $80~\gev$~\cite{no_tc}. The first set of new,
additional variables~\cite{heptop3} then included a combination of the
usual filtered top mass~\cite{bdrs} with a pruned top
mass~\cite{pruning}. In this upgrade we introduce a fat jet radius up
to $R=1.8$ for moderately boosted tops and allow for a choice of
Cambridge--Aachen~\cite{ca_algo} and $k_T$~\cite{kt_algo,fastjet} jet
algorithms in all internal clustering and filtering steps except for
the mass drop condition. This improves the tagging performance for
highly boosted tops~\cite{heptop3}.  Recently, the algorithm was
slightly changed to avoid background shaping~\cite{heptop4}. In the
same study we added a low-$p_T$ mode based on Fox--Wolfram
moments~\cite{fwm} to incorporate angular correlations, extending the
tagging coverage to $p_{T,t} = 150~\gev$.

In this paper we present a detailed study of the
\textsc{HEPTopTagger2}, collecting all previous modifications, as well
as a whole range of new features targeted at multivariate analyses and
statistical approaches to single events~\cite{qjets,telescoping}. The
main body of the paper will focus on $Z'$ searches, where final-state
jet radiation turns out to be the limiting factor of the original
tagger. After resolving the issue with final-state radiation we will
step by step improve the tagging algorithm by defining and including
additional kinematic information. Finally, we will compare the
multivariate tagging performance with the leading projections based on
event deconstruction~\cite{event_deco}.\bigskip

The main background in fully hadronic $Z' \to t\bar t$ searches is QCD
multi-jets production, which allows us to directly translate all our
findings into a performance study based on tagging $t\bar{t}$ pairs in
the Standard Model. We will show these results together with a review
of the complete \textsc{HEPTopTagger2} algorithm and the code
interface in the Appendix.

%%%%%%%%%%%%%%%%%%%%%%%%%%%%%%%%%%%%%%%%%%%%%%%%%%%%%%%%%%%%%%%%%%%%%%
\section{Resonance reconstruction}
\label{sec:resonance}

The key challenge of any top tagger is its broad range of
applications and the related optimization of the algorithms and codes.
For example, the \textsc{HEPTopTagger} was developed to solve the
combinatorial problems in $t\bar{t}H$ searches~\cite{heptop1}. The
first public tagging code was presented for supersymmetric top partner
searches in semi-leptonic top decays~\cite{heptop2}. Its proposed
applications include single top production to experimentally separate
the $s$-channel and $t$-channel production
processes~\cite{single_top}.  However, its experimental application
during the first LHC run was the search for heavy resonances decaying
to hadronic $t\bar{t}$ searches~\cite{atlas}. For such a resonance
search the kinematic top tagger in combination with a $b$-tag showed a
similar performance as the usual, approximate reconstruction of
semileptonic $t\bar{t}$ pairs. In this paper we will present a set of
improvements towards the \textsc{HEPTopTagger2} for a $Z'$ search at
the 13~TeV LHC. Many of these improvements can be applied to other LHC
processes, as will be discussed in the Appendix.\bigskip

In using all available information from a pair of boosted top quarks,
event deconstruction is currently giving the leading performance
estimates for heavy resonance searches~\cite{event_deco}. For the
analysis in the main body of this paper we will follow the analysis
framework of Ref.~\cite{event_deco}, to eventually allow for a
comparison in Sec.~\ref{sec:comparison}.  For the signal we therefore
use \textsc{Pythia8}~\cite{pythia} to generate $Z^\prime\to t\bar{t}$
events with $m_{Z^\prime}=1500~\gev$ at 13~TeV collider energy.
Assuming the same couplings as for the Standard Model $Z$-boson would
yield a width of $\Gamma(Z^\prime) =47~\gev$; to be consistent with
the assumed experimental resolution in Ref.~\cite{event_deco} we
increase the width to $65~\gev$ and only simulate the vector
couplings. However, we will see that this choice of the physical $Z'$
width does not affect our results which are based on the reconstructed
fat jet kinematics.  For the $Z'$ decay we assume a 100\% branching
ratio to top pairs. The two backgrounds are continuum $t\bar{t}$
production which we simulate assuming $p_{T,t} > 400~\gev$, and QCD
di-jet production, also requiring $p_{T,j} > 400~\gev$. Again, we rely
on \textsc{Pythia8}, keeping in mind that for the pure QCD background
our di-jet rate might not be a conservative estimate. All top quarks
are forced to decay hadronically. Our simulations for the main body 
of the paper include underlying
event but do not account for pile--up or detector effects, unless explicitly 
mentioned. For a
completely realistic study of the signal and background efficiencies
of the new \textsc{HEPTopTagger2} we will have to rely on upcoming
experimental studies. For our multivariate tagging analyses we
optimize the background rejection with respect to the pure QCD
background, because it is by far dominant.

%%%%%%%%%%%%%%%%%%%%%%%%%%%%%%%%%%%%%%%%%%%%%%%%%%%%%%%%%%%%%%%%%%%%%%
\subsection*{Decay kinematics}

On the analysis level we first select events with at least two fat
jets with
\begin{equation}
p_{T,\text{fat}} > 400~\gev 
\qqquad \text{and} \qquad 
|y_\text{fat}|<2.5 \; ,
\label{eq:fat_cuts}
\end{equation}
reconstructed using the C/A algorithm~\cite{ca_algo} with cone size
$R = 1.5$, as implemented in \textsc{FastJet}~\cite{fastjet}. We limit
ourselves to the two hardest fat jets in each event for the $Z'$
search. The corresponding cut flow is given in
Tab.~\ref{tab:cutflow}. Using the old default \textsc{HEPTopTagger}
setup~\cite{heptop2} we find a double top tagging efficiency of
$\eps_\text{2tags} = 14\%$ in the signal, as shown in
Tab.~\ref{tab:cutflow}. If we apply a fixed invariant mass window
$m_{tt} \in [1200, 1600]~\gev$ on the tagged and reconstructed top
quarks, the $Z'$ tagging efficiency is $\eps_{Z'} = 10.2\%$. For the
$t\bar{t}$ background we find mis-tagging probabilities of
$\eps_\text{2tags} = 13.7\%$ and $\eps_{Z'} = 3.3\%$. For the QCD
background sample the double mistag rates are
$\eps_\text{2tags} = 6.6\cdot 10^{-4}$ and
$\eps_{Z'} = 1.5\cdot10^{-4}$. The QCD jets background exceeds the
continuum top pair production by a factor five after all
cuts.\bigskip

%---------------------------------------------------------------------
\begin{table}[b!]
\centering
\begin{tabular}{l|r|r|r}
  \hline
  & $Z^\prime\to t\bar{t}$ & $t\bar{t}$ & QCD \\ \hline 
  generator level & $10^5$ & $10^5$ (1.76~pb)& $8 \cdot10^6$ (1.93~nb)\\ 
  $\geq$ 2 fat jets Eq.\eqref{eq:fat_cuts} & 69142 &
  85284 (1.50~pb)& $6.7\cdot10^6$ (1.62~nb)\\ 
  hardest 2 fat jets HTT[JHEP1010] tagged & 9679 & 11706 (0.21~pb)& 4426 (1.07~pb)\\ 
  $m_{tt} \in [1200, 1600]~\gev$ & 7031 & 2817 (0.05~pb) & 978 (0.24~pb)\\ \hline
\end{tabular}
\caption{Number of events and the corresponding \textsc{Pythia8} cross
  section used for our analysis. The efficiencies $\eps_{S,B}$ for a
  $Z'$ extraction are defined as the ratio of the last to the second
  line in this table.}
\label{tab:cutflow}
\end{table}
%---------------------------------------------------------------------

\begin{figure}[t]
  \centering
  \includegraphics[width=0.4\textwidth]{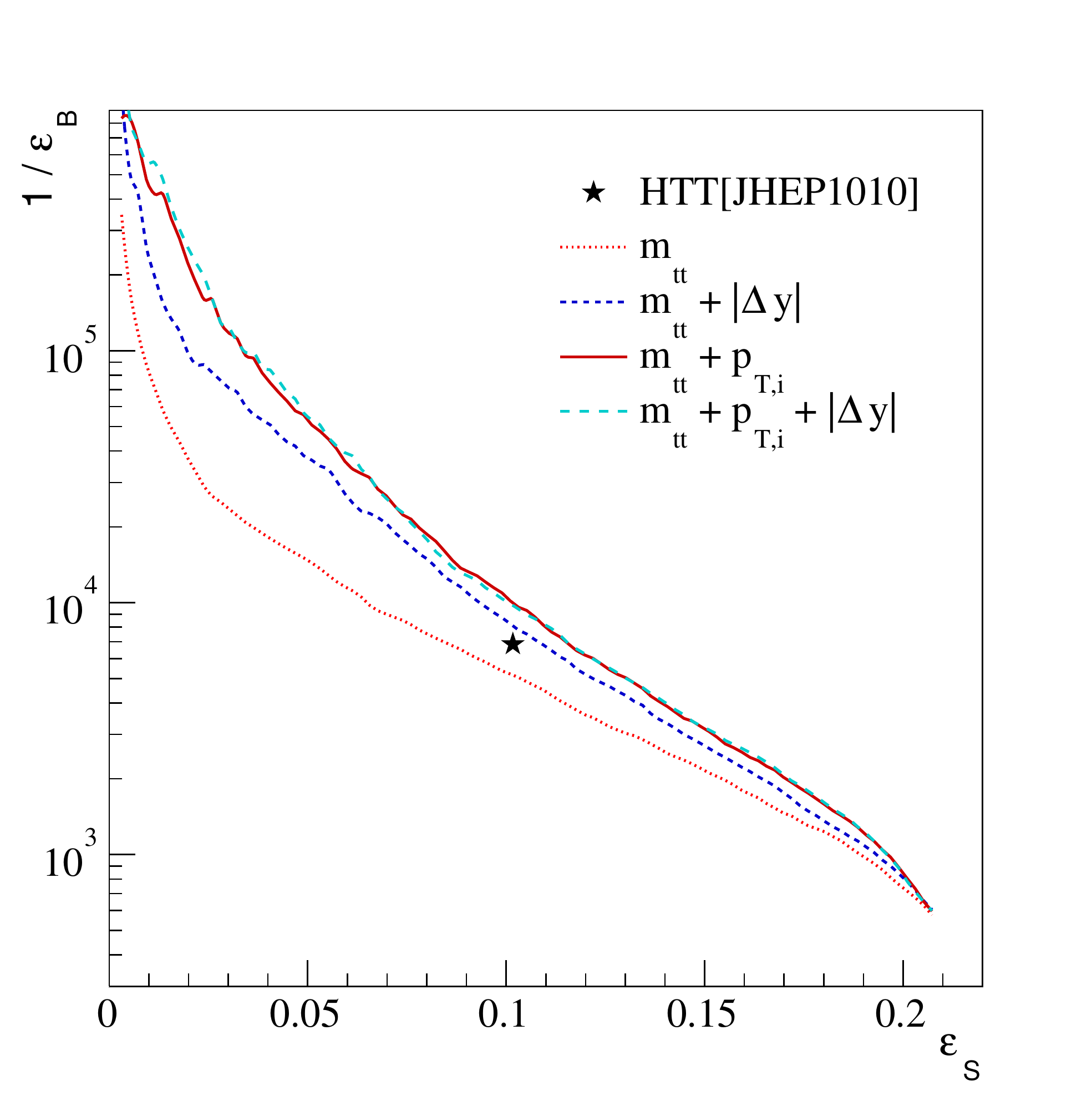}
  \hspace*{0.1\textwidth}
  \includegraphics[width=0.4\textwidth]{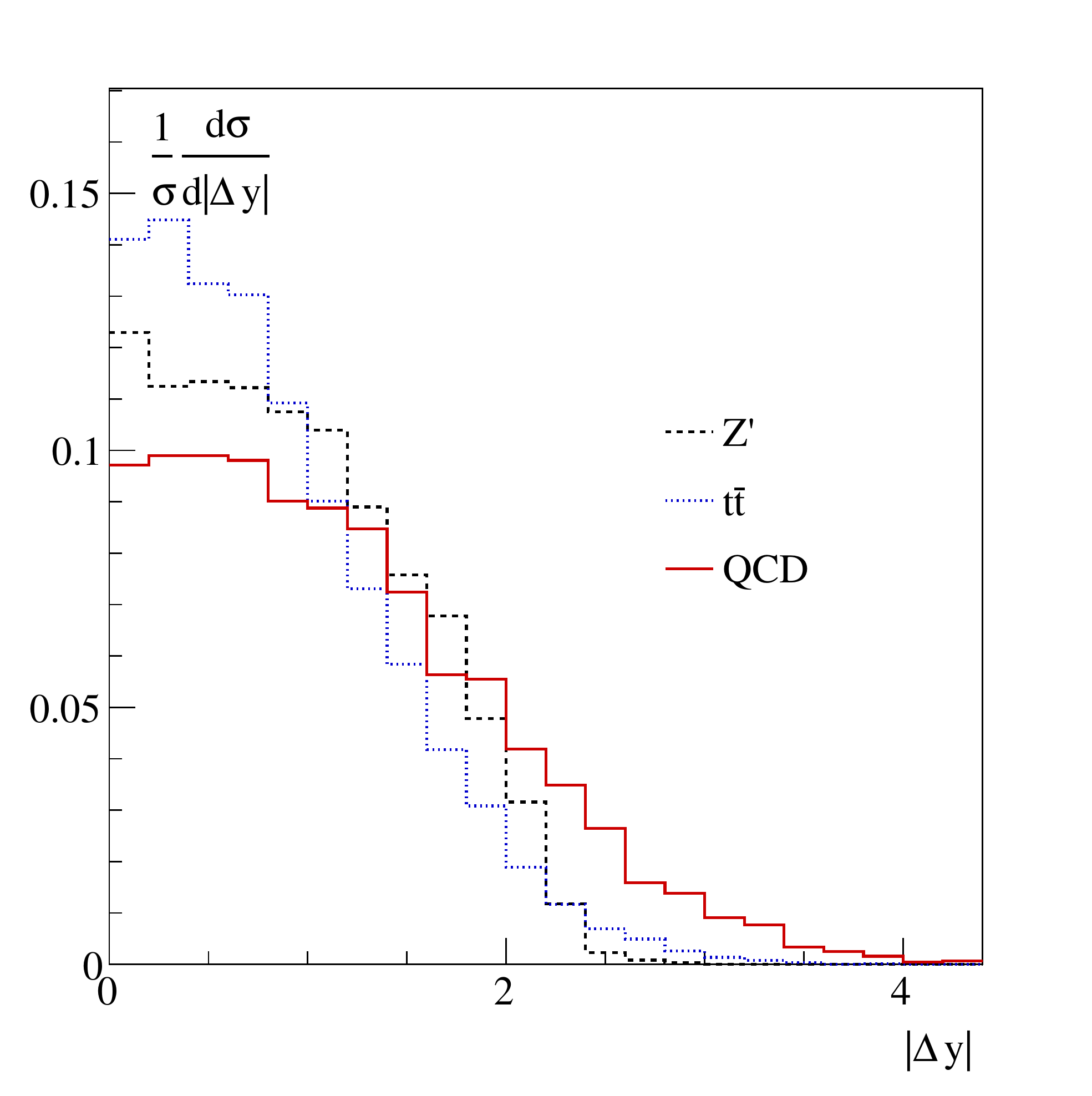}
  \caption{Left: ROC curves for the dominant QCD background vs. the
    $Z'$ signal after including additional kinematic information shown
    in Eq.\eqref{eq:vars_kin}. As in all figures the asterisk
    corresponds to the original \textsc{HEPTopTagger} described in
    Ref.~\cite{heptop2}.  Right: $|\Delta y|$ distribution of the
    reconstructed top quarks for signal and backgrounds.}
  \label{fig:dy}
\end{figure}

A straightforward improvement of the basic analysis shown in
Tab.~\ref{tab:cutflow} should be to replace the mass window by a
boosted decision tree (BDT) analysis, as implemented in
\textsc{Tmva}~\cite{tmva}, based on the reconstructed invariant mass
$m_{tt}$. In the left panel of Fig.~\ref{fig:dy} we first show the
results as receiver operator characteristic (ROC) curve, correlating
the best signal and background efficiencies based on a given set of
kinematic observables.  This approach has been used to improve and
benchmark the general performance of the
\textsc{HEPTopTagger}~\cite{heptop4}.  Because the QCD jet background
is dominant we always set up our multivariate analyses based on the
$Z^\prime$ signal and the QCD background sample. Compared to the
working point of the original public \textsc{HEPTopTagger}
tool~\cite{heptop2} with a fixed mass window
$m_{tt} \in [1200, 1600]~\gev$ the new \textsc{HEPTopTagger2}
including $m_{tt}$ in a multivariate analysis looks slightly
worse. The reason is the change in the order in the algorithm
described in the Appendix. It significantly reduces the background
sculpting, but at the expense of background rejection for example for
a constant signal efficiency. On the other hand, the reduced
background sculpting removes a major source of systematic uncertainty
when we need to interpret an $m_{tt}$ distribution which shows a peak
which could be due to a signal or to a sculpted background. Moreover,
it turns out that the difference between the old and new taggers
vanishes once both of them are used in a fully flexible multivariate
framework.\bigskip

%---------------------------------------------------------------------
%---------------------------------------------------------------------

For a better discrimination between signal and background we should
include additional variables in our multivariate analysis.  The
deterministic structure of the \textsc{HEPTopTagger} will still allow
for a particularly clear separation of the actual tagging and
reconstruction from a subsequent kinematic analysis based on the
reconstructed top momenta.  The first additional variable we include
is the rapidity difference between the two reconstructed top quarks,
$|\Delta y|$. The corresponding signal and background distributions
are shown in the right panel of Fig.~\ref{fig:dy}. While this variable
might not be too efficient in removing the $t\bar{t}$ continuum
background, events are visibly less central for QCD jets.  The
differences can hardly be translated into efficient kinematic cuts,
but they will help as part of a multivariate analysis. In the left
panel of Fig.~\ref{fig:dy} we show the corresponding improvement in
terms of ROC curves. In particular for low signal efficiencies $\eps_S
< 0.1$ we find a significant reduction of the background fake rates,
going beyond the working point of the first \textsc{HEPTopTagger}.

An obvious extension of our set of kinematic observables are the
transverse momenta of the reconstructed top quarks. Note that as part
of the ROC analysis we do not have to ensure that the different
kinematic variables are independent of each other, which would be
problematic for a combination of $m_{tt}$ and the $p_{T,t}$
distributions.  Again, the improvement from the transverse momentum
spectra is shown in the left panel of Fig.~\ref{fig:dy}. All this
illustrates that the kinematic
information on the tagged and reconstructed tops can increase the
background rejection by 50\% to 100\% for fixed signal tagging
efficiency. We also see that once we include the top--pair invariant
mass and the transverse momenta, the additional improvement from
$|\Delta y|$ vanishes, because the 2-particle final state is
essentially fully described. As kinematic observables in
our multivariate analysis we choose
\begin{equation} 
\{ \; m_{tt}, p_{T,t_1}, p_{T,t_2} \; \} 
\qqquad \text{(decay kinematics).}
\label{eq:vars_kin}
\end{equation}
%

%%%%%%%%%%%%%%%%%%%%%%%%%%%%%%%%%%%%%%%%%%%%%%%%%%%%%%%%%%%%%%%%%%%%%%
\subsection*{QCD jets}

In purely hadronic searches for new physics, QCD effects beyond fixed
order are a major issue in trying to theoretically understand the
signal and backgrounds. Before we devise strategies to deal with 
final-state radiation and initial-state radiation in heavy-resonance
searches we can estimate their effect on the naive tagger--based
analysis.

\begin{figure}[t]
  \centering
  \includegraphics[width=0.4\textwidth]{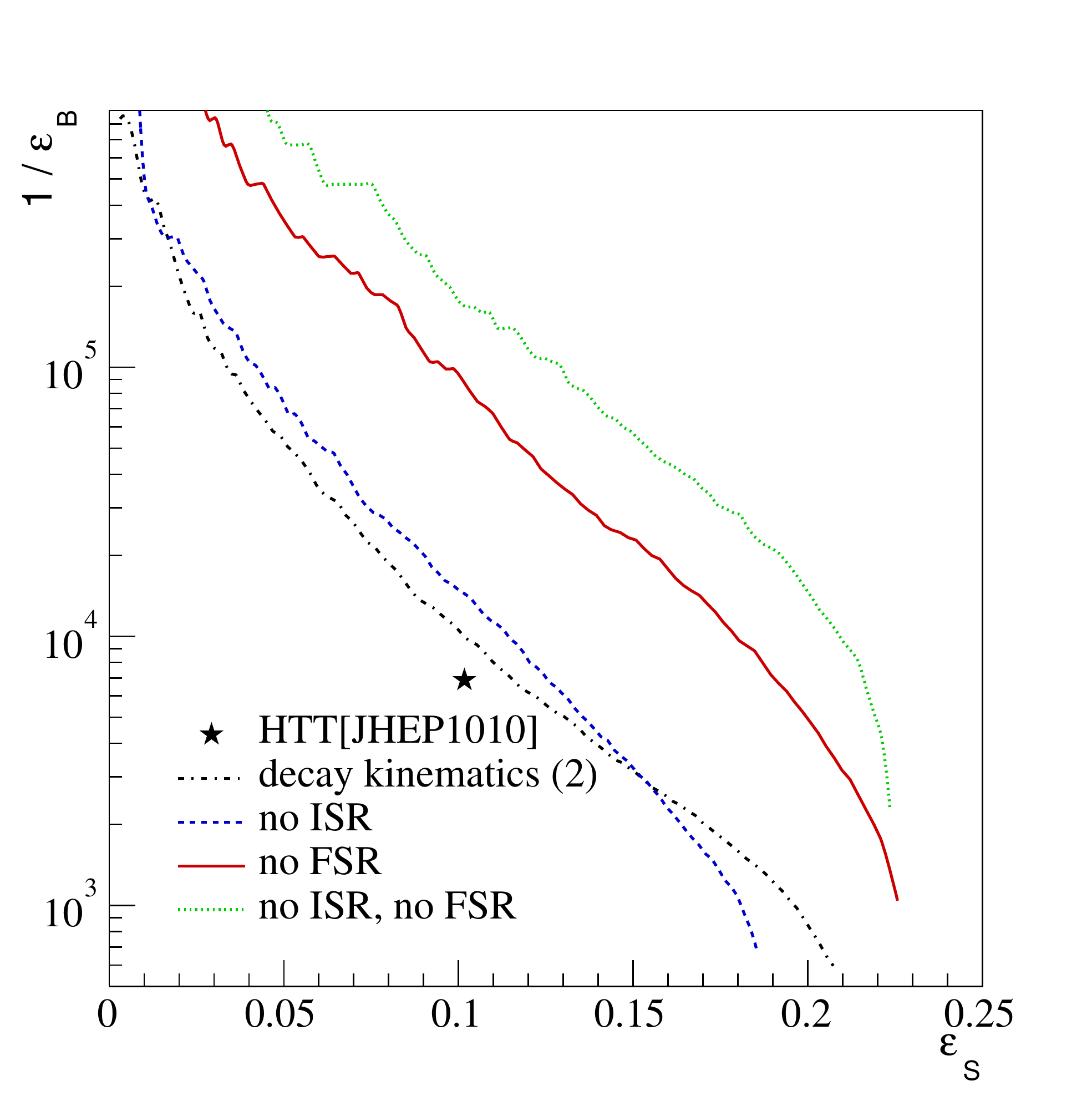}
  \caption{ROC curves for different combinations of initial-state jet
    radiation (ISR) and final-state jet radiation (FSR) in the $Z'$
    signal generation. The background is QCD with ISR and FSR for all
    curves.}
\label{fig:roc_xsr}
\end{figure}

On the Monte Carlo level it is possible to separately remove
initial-state radiation and final-state radiation from all signal
events. For 
the QCD jets background this is not sensible, because we need both
mechanisms to generate a sufficient jet multiplicity.  The ROC curves
in Fig.~\ref{fig:roc_xsr} show the expected improvements in the
absence of additional signal jets. We see that the leading effect
spoiling the signal extraction is final-state radiation (FSR). 
Initial-state radiation (ISR) affects top tagging in two ways. First, the
additional QCD jets can mimic for example the softer $W$-decay jet and
degrade the tagging efficiency through combinatorics. On the other
hand, ISR jets recoil against the $Z'$, affecting the $p_T$ spectrum
of the top quarks. In particular the tagging of the softer top decay
can benefit from this recoil, which means that for large signal
efficiency the results without ISR become worse than those with all
jet activity included.

As a whole, the results shown in Fig.~\ref{fig:roc_xsr} indicate 
potentially significant improvements of top taggers when we target the
different effects of QCD jet radiation. We will show in the following
subsection how a deterministic top tagger is limited by final-state
radiation and how the new \textsc{HEPTopTagger2} can avoid these
issues. Combinatorial problems related to initial-state radiation will
then be one of the key topics in Sec.~\ref{sec:optr}.

%---------------------------------------------------------------------
%---------------------------------------------------------------------

%%%%%%%%%%%%%%%%%%%%%%%%%%%%%%%%%%%%%%%%%%%%%%%%%%%%%%%%%%%%%%%%%%%%%%
\subsection*{Final-state radiation}

Final-state radiation (FSR) turns one of the key advantages of our top
tagger into a significant problem: unlike some other top tagging
approaches, the \textsc{HEPTopTagger} returns the 4-momentum of the
tagged top, including a cut on the reconstructed top mass
$m_\text{rec} \in [150,200]~\gev$~\cite{heptop4}.  This allows us to
trivially reconstruct $m_{Z'}$. Final-state radiation off the top
decay products will be captured by the jet clustering and contribute
to the correct filtered top mass value~\cite{bdrs}.  This way it will
not pose a problem as long as the $Z'$ decays to on-shell tops.

\begin{figure}[t]
  \includegraphics[width=0.4\textwidth]{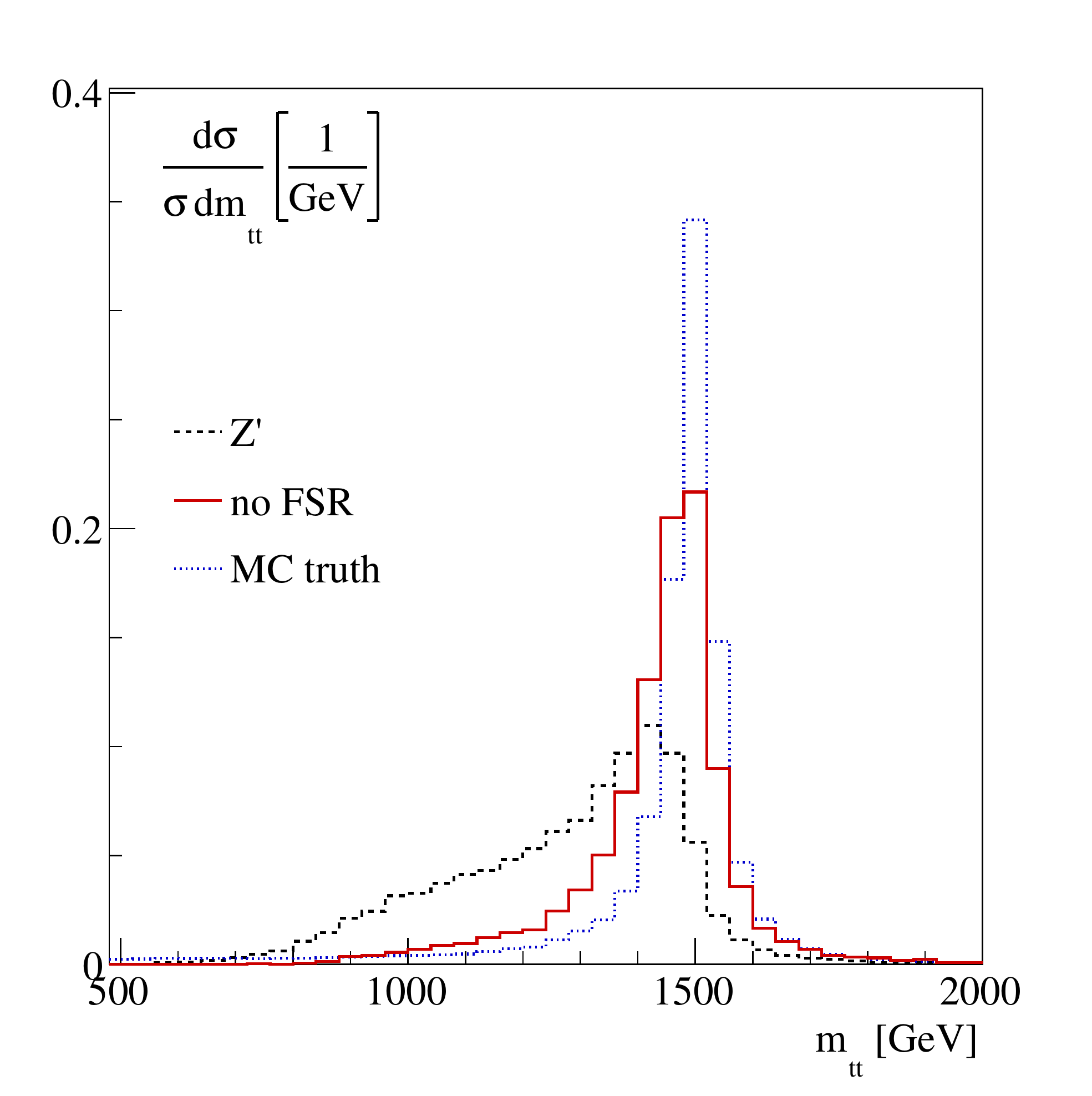}
  \hspace*{0.1\textwidth}
  \includegraphics[width=0.4\textwidth]{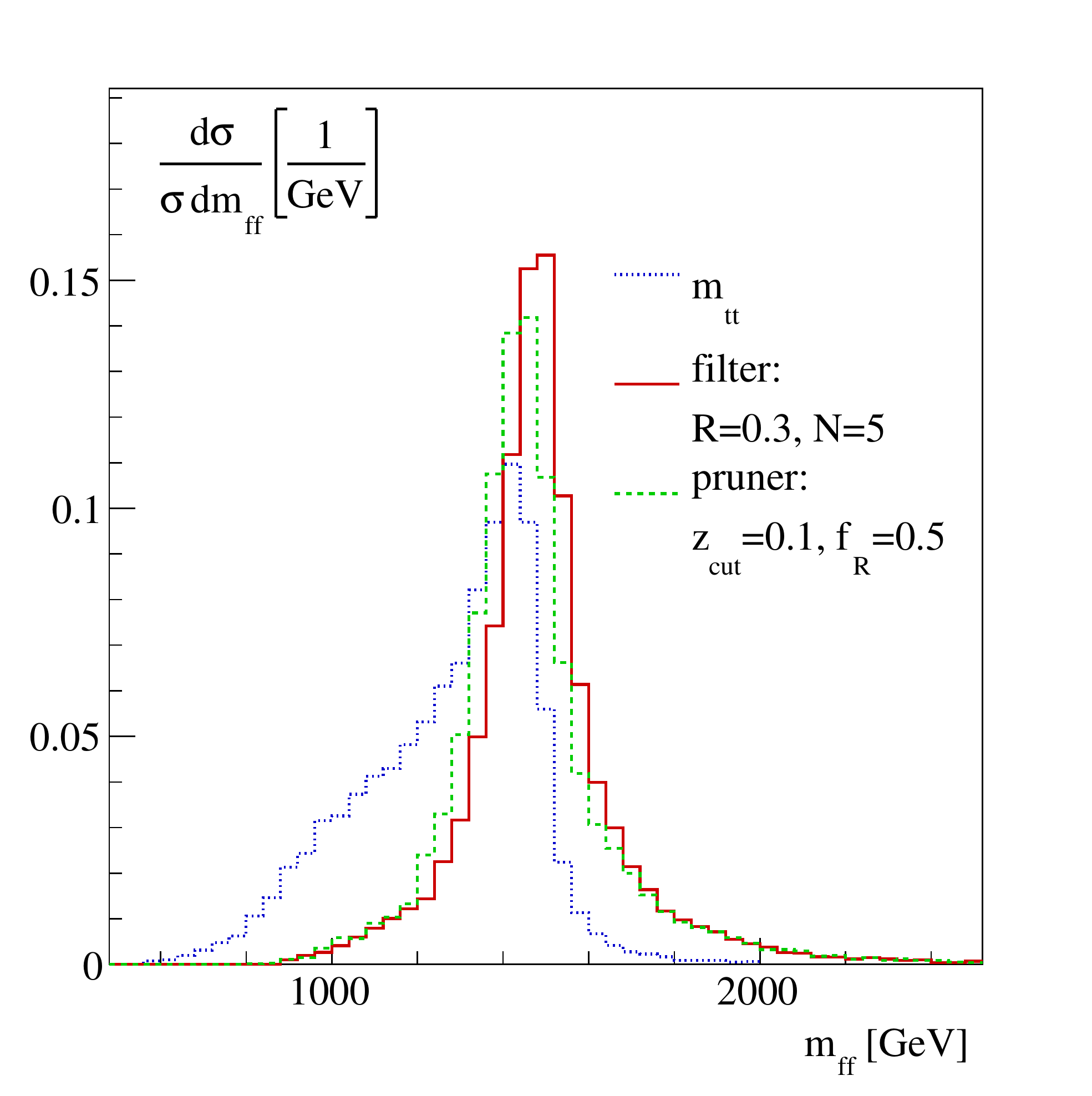}
  \caption{Effect of final-state radiation on the invariant mass of
    the tagged and reconstructed $t\bar{t}$ system $m_{tt}$ for the
    $Z'$ signal (left) and different approaches to reconstruct the $Z'$
    mass peak (right). Monte Carlo truth is $\sqrt{p_{Z'}^2}$ with an
    assumed width of $65~\gev$.}
  \label{fig:mff}
\end{figure}

However, if the $Z'$ decays to slightly off-shell tops, which turn
themselves into on-shell tops, this final-state radiation off the
intermediate top mis-aligns the actual $Z'$ with the $Z'$ as
reconstructed from the top quarks at the moment they decay.  Because
the hard radiated gluon does not enter the top reconstruction, the top
tag will pass, but lead to an underestimated $m_{Z'}$ value. In the
left panel of Fig.~\ref{fig:mff} we indeed see that the $m_{tt}$
distribution for the top-tagged signal correctly peaks around
$m_{Z'}$, but develops a sizeable asymmetric tail towards smaller
$m_{tt}$ values.  While the details of this asymmetric tail from
\textsc{Pythia8} should be subject to a detailed Monte Carlo study, we
simply confirm that turning off final-state radiation by hand gets rid
of it almost entirely.  The remaining slight broadening as well as a
minimal tail towards smaller $m_{tt}$ values is due to small losses in
the top 4-momentum reconstruction of the tagger.  At higher values of
$m_{Z'}$ the asymmetric tail is further enhanced.\bigskip

%---------------------------------------------------------------------
%---------------------------------------------------------------------

The problem with large asymmetric tails from final-state radiation is
that they cannot simply be corrected for in a universal top
tagger. The basic structure of the \textsc{HEPTopTagger} has to
identify and reconstruct top quarks, rather than the decay products of
a heavy $Z'$ resonance. Therefore, we do not modify the actual tagger,
but we account for final-state radiation through an additional set of
kinematic observables.

%---------------------------------------------------------------------
\begin{table}[!b]
\centering
\begin{tabular}{l|r|r|r|r|r}
\hline
  & $m_\text{peak}$~[GeV]& $\Gamma$ [GeV] & $\eps^{\pm
    150}_{Z'}$& $1/\eps^{\pm 150}_{tt}$ & $1/\eps^{\pm
    150}_\text{QCD}$\\ \hline
  $m_{tt} \in [1200,1600]~\gev$ & -- & -- & 0.136 & 22 & 2805 \\ \hline
  unfiltered     & 1539 & 167 & 0.141 & 21 & 1960 \\ \hline
  $R=0.3$, $N=4$ & 1457 & 152 & 0.146 & 28 & 2218 \\ 
  $R=0.3$, $N=5$ & 1477 & 144 & 0.150 & 25 & 2098 \\
  $R=0.3$, $N=6$ & 1489 & 139 & 0.151 & 25 & 2052 \\
  $R=0.3$, $N=7$ & 1496 & 144 & 0.151 & 24 & 2043 \\ \hline
  $R=0.2$, $N=5$ & 1443 & 140 & 0.141 & 29 & 2329 \\
  $R=0.3$, $N=5$ & 1477 & 144 & 0.150 & 25 & 2098 \\
  $R=0.4$, $N=5$ & 1500 & 144 & 0.151 & 24 & 2030 \\
  $R=0.5$, $N=5$ & 1515 & 143 & 0.148 & 23 & 1993 \\ \hline
  pruning $z=0.1$, $f_R=0.5$ & 1443 & 150 & 0.138 & 26 & 2075 \\ \hline
\end{tabular}
\caption{Breit--Wigner fits and performance of different grooming
  approaches. The quoted efficiencies are based on a window for the
  invariant mass of the two filtered fat jets $|m_\text{ff}-m_{Z'}|<150~\gev$.}
\label{tab:fit}
\end{table}
%---------------------------------------------------------------------

Following the brief discussion above, including the kinematics of the
fat jet in addition to the reconstructed top 4-momentum should remove
the broad asymmetric tail in the reconstructed $m_{Z'}$ values.
Again, we first select events with two tagged tops, including the top
mass condition.  Instead of using the 4-momenta of the tagged tops, we
now reconstruct the $Z'$ mass from the 4-momenta of the two fat jets
of size $R=1.5$, which eventually lead to the top tags. In the
presence of underlying event and initial-state radiation the naive
$m_\text{ff}$ distribution peaks roughly at the correct $Z'$ mass and
shows symmetric tails.  To use the invariant mass of the two fat jets
we need to apply filtering~\cite{bdrs}.  In the right panel of
Fig.~\ref{fig:mff} we compare the filtered invariant mass from the two
fat jets~\cite{bdrs} and its pruned value~\cite{pruning}, both as
implemented in \textsc{FastJet}~\cite{fastjet}. As a reference we also
show the $m_{tt}$ distribution from the left panel of the same
figure. Unlike the reconstructed $m_{tt}$ distribution, both, the
filtered and the pruned $m_\text{ff}$ distributions give symmetric
peaks around the correct $m_{Z'}$ value.\bigskip

To be able to use the filtered $m_\text{ff}$ values in our
\textsc{HEPTopTagger} analysis we confirm that filtering and
pruning give stable numerical results for the invariant mass of the
two fat jets.  Results for different parameter settings are listed in
Tab.~\ref{tab:fit}. We give the peak positions, which would be subject
to a proper calibration, the fitted Breit--Wigner widths for the
symmetric peaks, and the tagging performances for a fixed mass window
$|m_\text{ff}-m_{Z'}| < 150~\gev$. Replacing the Breit--Wigner width
with a Gaussian would make no difference, but give a poorer modelling
of the tails. Typical widths of the reconstructed $Z'$ mass peak will
range around $145~\gev$, roughly twice the assumed particle width of
$65~\gev$. Even in the absence of detector effects, this resolution
will replace the assumed particle width of $65~\gev$ in all of the
following analysis.  The constant numbers in Tab.~\ref{tab:fit}
confirm that the $m_\text{ff}$ criterion is stable for different
filtering parameters as well as pruning.

\begin{figure}[t]
  \includegraphics[width=0.4\textwidth]{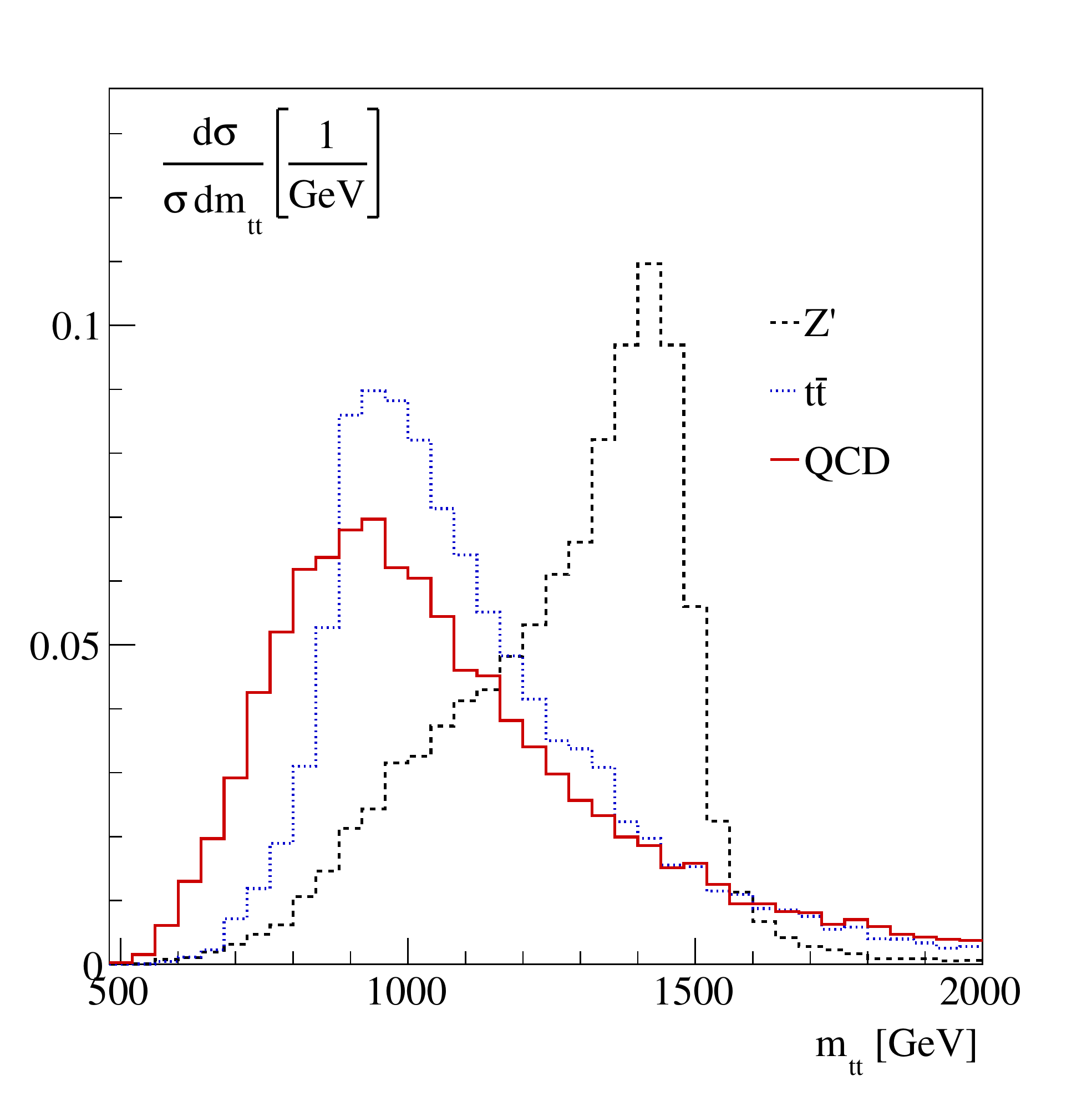}
  \hspace*{0.1\textwidth}
  \includegraphics[width=0.4\textwidth]{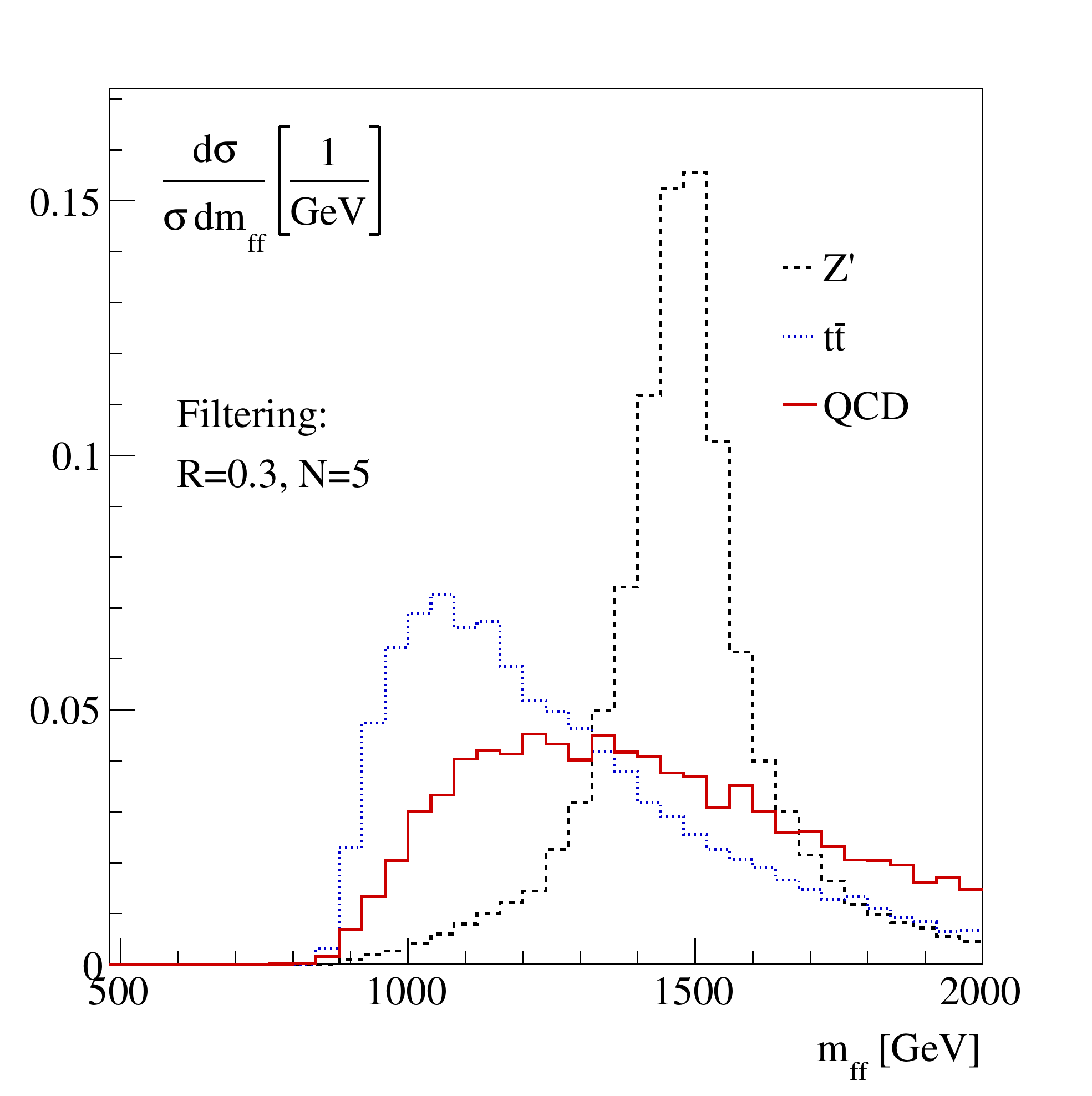}
  \caption{Reconstructed mass distribution of the $Z'$ signal and the
    backgrounds based on the tagged tops (left) and the corresponding
    filtered fat jets (right).}
  \label{fig:masses}
\end{figure}

On the other hand, the results shown in Tab.~\ref{tab:fit} also
indicate that simply replacing the $m_{tt}$ window by a filtered
$m_\text{ff}$ value will not improve the $Z'$ extraction.  In
Fig.~\ref{fig:masses} we show that the steeply falling QCD jets
background now has a maximum around $m_\text{ff} = 1.3~\tev$, while
for the reconstructed top quarks there exists a much more pronounced
maximum around $m_{tt} = 900~\gev$. The reason is that top tagging
removes events with many hard QCD jets in two steps: first requiring
the correct top mass value from three assumed top decay products, and
second when applying the $Z'$ mass window. If we remove the first
step, the second one has to deal with larger backgrounds at high
$m_\text{ff}$ values.

%---------------------------------------------------------------------
%---------------------------------------------------------------------

\begin{figure}[t]
  \centering
  \includegraphics[width=0.4\textwidth]{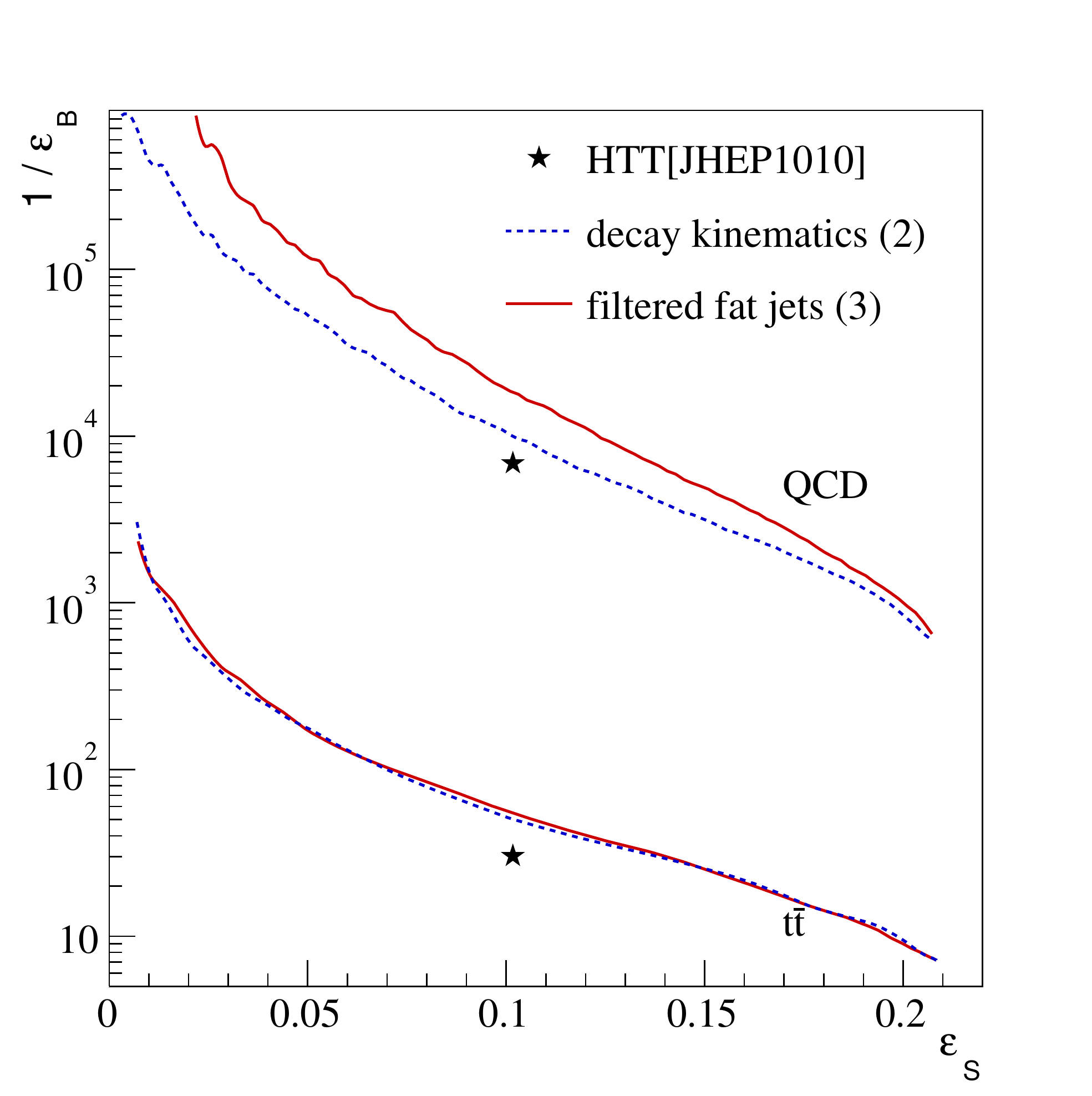}
  \hspace*{0.1\textwidth}
  \includegraphics[width=0.4\textwidth]{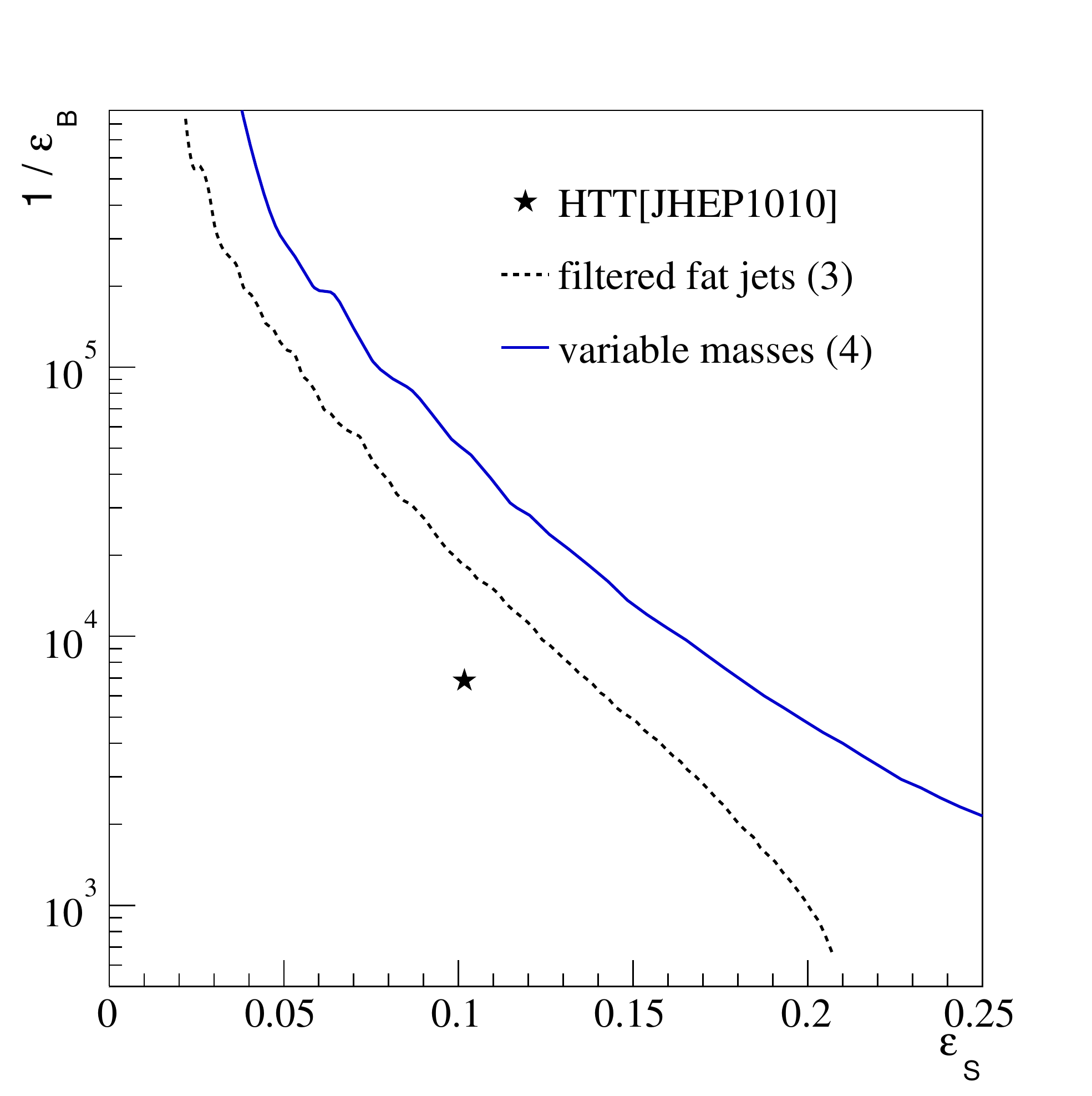}
  \caption{Left: performance of the multivariate analysis including
    the information on the fat jet, as given in
    Eq.\eqref{eq:vars_kin}, Eq.\eqref{eq:vars_fat} and
    Eq.\eqref{eq:vars_window}. Only in this plot do we optimize for
    $t\bar{t}$ and QCD backgrounds separately. Right: performance
    curve for the full analysis only accounting for the dominant QCD
    jet background.}
  \label{fig:roc_wp}
\end{figure}

If we want to include final-state radiation and at the same time
benefit from its additional information, we need to keep $m_\text{ff}$
as well as $m_{tt}$ in our analysis, and not apply a simple mass
window on the $m_{tt}$ distribution.  The kinematics of the $Z'$-decay
is then described by
\begin{equation} 
\{ \; m_{tt}, m_\text{ff}, p_{T,t_1}, p_{T,t_2}, p_{T,\text{f}_1}, p_{T,\text{f}_2} \; \} 
\qqquad \text{(filtered fat jets).}
\label{eq:vars_fat}
\end{equation}
All default settings of the \textsc{HEPTopTagger} are listed in the Appendix.
We filter the fat jets using $R=0.3$ and keep the $N=5$ hardest
substructures.  In the left panel of Fig.~\ref{fig:roc_wp} we show the
corresponding ROC curves. Unlike in the rest of the paper we study the
$t\bar{t}$ and QCD jets backgrounds separately. The improvement of
the full multivariate tagger including the fat jet information of
Eq.\eqref{eq:vars_fat} is obvious for both backgrounds.  In the right
panel of Fig.~\ref{fig:roc_wp} we first show the same improvement, but
using a BDT trained on the QCD jets background only.  Compared to the
original \textsc{HEPTopTagger} we achieve an improvement of up to a
factor 2 in $1/\eps_B$ for constant signal efficiency. We note that
for the QCD background the combination of mis-tagged top
kinematics and fat jet kinematics goes beyond the description of the
hard process. For example initial-state radiation, sensitive to the
color structure of the signal and the background, will be captured in
this combination of observables. On the other hand, because the fat
jets are defined using the standard jet algorithms and show a stable
filtering performance, we do not envision major experimental problems
provided pile-up subtraction works as well as expected.\bigskip

%---------------------------------------------------------------------
%---------------------------------------------------------------------

The set of kinematic observables listed in Eq.\eqref{eq:vars_fat}
still relies on the deterministic \textsc{HEPToptagger} output. This
means that the identification of a $Z'$ signal event is limited by the
efficiency of two top tags. The choice of a working point in the top
tagging algorithm will therefore limit our over-all efficiency. On the
other hand, we already know that for hadronic $Z'$ searches the QCD
jets background is dominant and will only be reduced through a
combination of top tags and $Z'$ mass reconstruction. 

In addition, we omit a fixed mass window for the reconstructed top
mass $m_\text{rec}$. Instead, we widely open the top mass and $W$-mass
constraints in the tagging algorithm. For each of the tops the
corresponding $m_\text{rec}$ value then becomes an output of the
tagger. We provide the multivariate $Z'$ analysis with the smaller and
larger of these two output $m_\text{rec}$ values, which we label as
$m_\text{rec}^\text{min}$ and $m_\text{rec}^\text{max}$
respectively. Similarly, we avoid a fixed window for the ratio of the
$W$-mass to the top mass, parametrized as $f_W$ in the tagging
algorithm. Its deviation from the true value is given by the value of
$f_\text{rec}$ defined in the Appendix. In the multivariate analysis
we include the maximum of the two $f_\text{rec}$ values corresponding
to each tagged top.
\begin{equation} 
\{ \; m_{tt}, m_\text{ff}, p_{T,t_1}, p_{T,t_2}, p_{T,\text{f}_1}, p_{T,\text{f}_2},
m_\text{rec}^\text{min}, m_\text{rec}^\text{max}, f_\text{rec}^\text{max} \; \} 
\qqquad \text{(variable masses).}
\label{eq:vars_window}
\end{equation}
The result is shown in the right panel of Fig.~\ref{fig:roc_wp}, where
the range of accessible efficiencies eventually extends to $56\%$.
Altogether, the analysis based on the set of kinematic variables shown
in Eq.\eqref{eq:vars_window} gives us an improvement of up to a factor
5 in background rejection for a constant $Z'$-signal efficiency.

%%%%%%%%%%%%%%%%%%%%%%%%%%%%%%%%%%%%%%%%%%%%%%%%%%%%%%%%%%%%%%%%%%%%%%
\section{Updated tagger}
\label{sec:optr}

Fat jets with a geometric size of $R=1.5$ or even $R=1.8$ have shown
to be powerful new analysis objects at the LHC. The radius of the fat
jet is directly related to the energy or boost of the heavy particles
which can be captured. This means that a multi-purpose top tagger will
be based on as large fat jets as possible. However, to realize their
potential such large jets require additional treatment linked to their
large geometric size. Without a dedicated analysis step, underlying
event and pile-up will almost entirely wash out any structure inside
the fat jet. Filtering~\cite{bdrs} as an integral part of all versions
of the \textsc{HEPTopTagger}~\cite{heptop1,heptop2} effectively
reduces the geometric size of the fat jet used to reconstruct the top
4-momentum by introducing a second clustering stage with higher
resolution. This solves the problem with underlying event and
pile-up, but there remains a combinatorial problem caused for example by
initial-state radiation. In particular the softer of the two subjets
from the $W$-decay can easily be faked by a typical QCD jet inside the
fat jet. This will lead to a wrong reconstruction of the top
4-momentum, which we can only counter by applying harder tagging
requirements and hence reducing the tagging efficiency. These
so-called type-2 tags~\cite{heptop3}, where only two of three top
decay jets can be identified with a parton-level decay quark have been
in the focus of \textsc{HEPTopTagger} studies at moderate
boost~\cite{heptop3,heptop4,moderate}. In the reconstruction of heavy
resonances we can solve the problem of (too) large fat jets by
adapting the size of the fat jet to the kinematics of the tagged
top. It turns out that this adaptive size of the fat jet also gives us another
powerful kinematic variable for the multivariate analysis. Finally, we
will show how this optimalR modification of our tagging algorithm can be
further improved by including $N$-subjettiness variables.

%%%%%%%%%%%%%%%%%%%%%%%%%%%%%%%%%%%%%%%%%%%%%%%%%%%%%%%%%%%%%%%%%%%%%%
\subsection*{OptimalR mode}

There have been different attempts to adjust the size of the fat jet
for example based on the transverse momentum of the fat
jet~\cite{hopkins,telescoping,variableR}, but none them lead to a
dramatic effect in the performance of taggers. We instead choose a
purely algorithmic way of determining the minimum size of the fat
jet~\cite{strebler}. Assuming that three top decay jets are captured
by the fat jet we can run the standard \textsc{HEPToptagger} algorithm
to determine the top mass from the three leading
subjets~\cite{heptop4}. For a large fat jet size, typically $R=1.5$ or
$R=1.8$, we compute a reference value of $m_\text{rec}$, which should
be around the top mass. In the usual tagging algorithm, this
computation of $m_\text{rec}$ from filtered subjets takes into account
final-state radiation off the on-shell top.  We then reduce the size
of the fat jet in steps of $\Delta R = 0.1$ and compute the
corresponding values of $m_\text{rec}(R)$. In case of several possible
triplets, this includes the step of choosing the one closest to the
physical top mass, as described in step~(5) in the Appendix.  As a
function of the decreasing jet size $R$ the fat jet mass
$m_\text{rec}(R)$ will form a stable plateau, until the reduced fat
jet will be too small to capture all three top decay jets. At this
point $m_\text{rec}(R)$ will leave the plateau and show a significant
drop. For $R=1.5$, which is sufficient for the $Z'$ mass in our study,
we define this drop through
\begin{equation}
\frac{m_\text{rec}^{(1.5)} - m_\text{rec}(R)}{m_\text{rec}^{(1.5)}} > 0.2
\qquad \Leftrightarrow \qquad 
R < R_\text{opt} \, .
\label{eq:r_drop} 
\end{equation}
Once the shrinking fat jet passes this condition we go back one step
to the last $R$ value on the plateau and define this value as
$R_\text{opt}$. The smallest value we allow in this study is
$R_\text{opt}=0.5$, but for $p_{T,t} \gtrsim 1$~TeV this value can be
adjusted in the tagger setup.  This value could be a challenge of the
calorimeter resolution, so the corresponding results are subject to
tests based on a full detector simulation in ATLAS and in CMS. In this
paper we typically arrive around $R_\text{opt} = 0.6$.  The tagging
result for this $R_\text{opt}$ value will be the output of the top
tagger.\bigskip

Measuring $R_\text{opt}$ defines another useful variable for the top
tagger, because we can also predict $R_\text{opt}$ from the fat jet
kinematics. A similar reasoning is used in the original
\textsc{HEPTopTagger} algorithm, where a consistency condition on the
reconstructed top momentum $p_{T,t} > 200~\gev$ ensures that the
reconstructed top can actually be captured in the fat jet. In the
optimalR mode we first determine the transverse momentum of the filtered
fat jet, $p_{T,\text{f}}$ as described in the previous
section. Including up to ten hardest subjets after a filtering step
with $R_\text{filt} = 0.2$ turns out to give the best estimate of
$p_{T,\text{f}}$ for this purpose. Reducing this number to five
subjets has no measurable effect on the width of the reconstructed
$p_{T,\text{f}}$ distribution, but slightly shifts its maximum to
smaller values~\cite{strebler}. The final number will be subject to an
independent optimization in ATLAS and CMS. 

\begin{figure}[t]
  \includegraphics[width=0.4\textwidth]{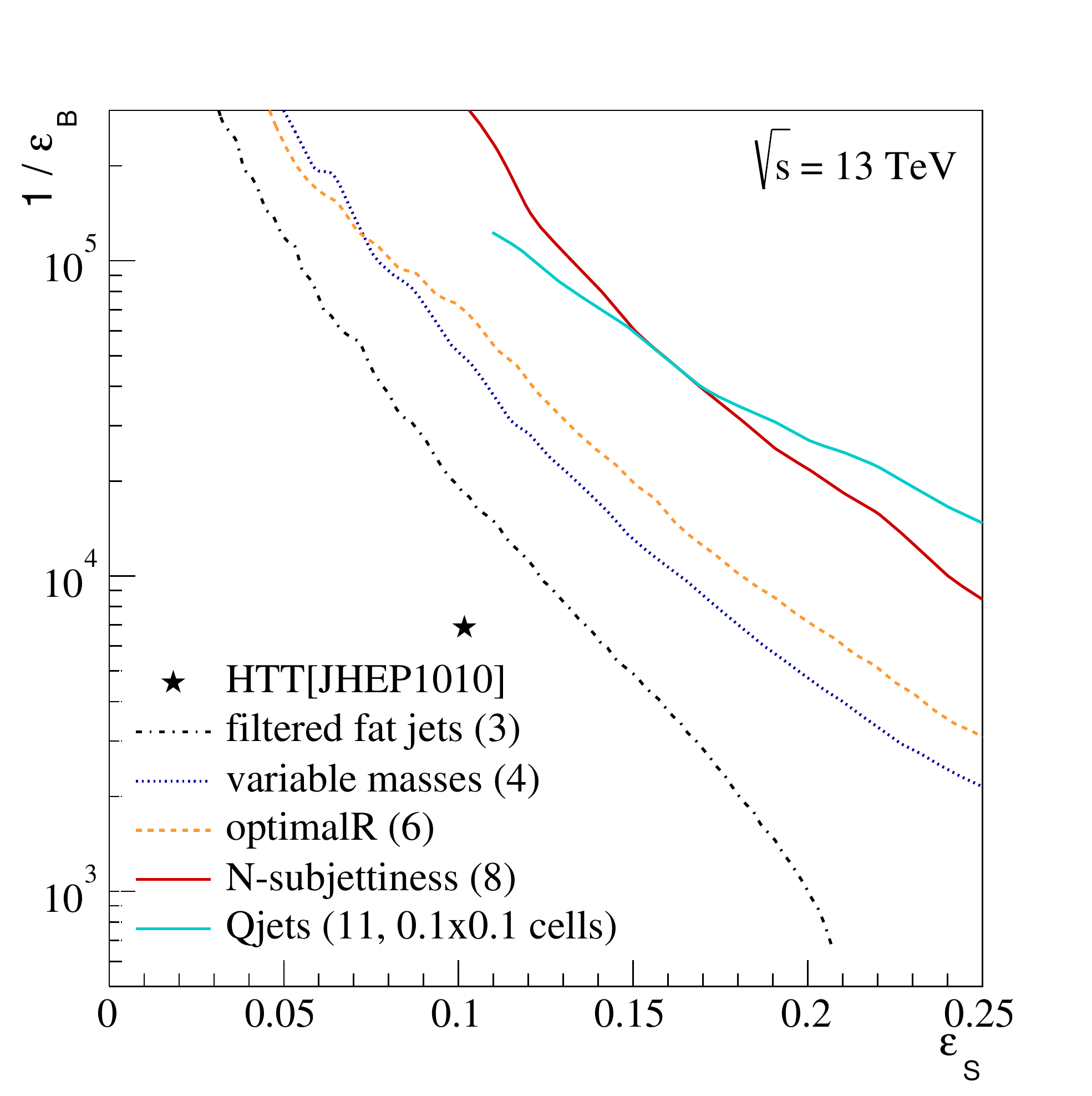}
  \hspace*{0.1\textwidth}
  \caption{Performance of the optimalR mode based on the kinematic
    variables in Eq.\eqref{eq:vars_optr}, including $N$-subjettiness
    variables as defined in Eq.\eqref{eq:vars_nsub}, and including
    \textsc{Qjets}. As described in the text, for \textsc{Qjets} we
    need to require a finite calorimeter resolution, while all other
    curves do not include any detector effects. We only consider the
    dominant QCD background.}
\label{fig:roc_new}
\end{figure}

For $p_{T,\text{f}} > 200~\gev$ we derive a closed form by fitting a
function $R_\text{opt}^\text{(calc)} \propto 1/p_{T,\text{f}}$ to
simulated data, as described in the Appendix. The kinematic variables
in our the multivariate tagger now read
\begin{equation} 
\{ \; m_{tt}, m_\text{ff}, p_{T,t_1}, p_{T,t_2}, 
      p_{T,\text{f}_1}, p_{T,\text{f}_2}, 
      m_\text{rec}^\text{min}, m_\text{rec}^\text{max}, f_\text{rec}^\text{max},
      R_\text{opt} - R_\text{opt}^\text{(calc)} \; \} 
\qqquad \text{(optimalR).}
\label{eq:vars_optr}
\end{equation}
For this case of two top tags we choose $R_\text{opt} -
R_\text{opt}^\text{(calc)}$ as the maximum deviation of the tagged
tops. In this form all subsequent kinematic variables linked to the
top tags will be evaluated with the fat jet size $R_\text{opt}$. For
the $Z'$ search $R_\text{opt}^\text{(calc)}$ will be strongly
correlated with other kinematic variables listed in
Eq.\eqref{eq:vars_optr}. We nevertheless include it in the BDT because
the general multivariate \textsc{HEPTopTagger2} described in the
Appendix will not include the top momenta in the tagging. The increase
of the tagging performance from the optimalR mode is shown in the left
panel of Fig.~\ref{fig:roc_new}. While for small signal efficiencies
the curves for optimalR and for the variable mass setup of
Eq.\eqref{eq:vars_window} are identical within numerical fluctuations,
we observe a significant improvement for larger signal efficiencies.

%---------------------------------------------------------------------
%---------------------------------------------------------------------

%%%%%%%%%%%%%%%%%%%%%%%%%%%%%%%%%%%%%%%%%%%%%%%%%%%%%%%%%%%%%%%%%%%%%%
\subsection*{N-subjettiness}

The arguably simplest question we can ask as part of a top tagger is
the number of hard subjets inside the fat jet with a given jet
mass. This number of subjets can be defined through an observable
similar to event shapes like for example thrust, called
$N$-subjettiness~\cite{nsubjettiness,better_obs}. It is based on $N$ reference
axes which are required to match the $k$ hard substructures,
\begin{equation}
 \tau_N = \frac{1}{R_0 \sum_k p_{T,k}} \sum_k p_{T,k} \min \left(
   \Delta R _{1,k}, \Delta R _{2,k}, \cdots, \Delta R _{N,k} \right) \;,
 \label{eq:tau_N}
\end{equation}
where $\Delta R_{i,j}$ is the geometric separation between the axis
$i$ and the substructure $k$. In this form $N$-subjettiness
parametrizes the deviation of the energy flow away from $N$ jets not
only related to an integer number of subjets, but also reflecting the
color structure and the related radiation pattern.

In terms of original definition~\cite{nsubjettiness} we fix the
exponent to $\beta = 1$. $R_0$ is an intrinsic cone size, chosen such
that $\tau_N < 1$.  Small values of $\tau_N \to 0$ indicate that the
complete substructure is described by $N$ axes, indicating that there
are at most $N$ relevant substructures.  The ratio $\tau_N/\tau_{N-1}$
will therefore become small for a fat jet with $N$ hard subjets.  For
top tagging the ratio $\tau_3/\tau_2$ will be most useful and can even
be used as a tagger itself. Higher $\tau_N$ values will contribute to
a multivariate analysis of $N$-subjettiness, describing the jet
radiation pattern around the assumed three partonic top decay
momenta.\bigskip

We will use $N$-subjettiness as an additional variable in our
multivariate \textsc{HEPTopTagger}. Originally, this combination did
not lead to a significant improvement when added to the A-shaped
cuts~\cite{heptop4}. However, when we open the cut $f_W$ on the
reconstructed ratio $m_W/m_t$ we observe a significant
improvement for the extended set of kinematic variables.  The complete
set of relevant kinematic variables, now including $N$-subjettiness
variables before and after filtering, is
\begin{equation} 
\{ \; m_{tt}, m_\text{ff}, p_{T,t_1}, p_{T,t_2}, 
      p_{T,\text{f}_1}, p_{T,\text{f}_2}, 
      m_\text{rec}^\text{min}, m_\text{rec}^\text{max}, f_\text{rec}^\text{max},
      R_\text{opt} - R_\text{opt}^\text{(calc)}, 
      \tau_{1,N}, \tau_{1,N}^\text{(filt)},\tau_{2,N}, \tau_{2,N}^\text{(filt)} \; \} 
\quad \text{($N$-subjettiness).}
\label{eq:vars_nsub}
\end{equation}
For more details on the $N$-subjettiness variables we refer to the
Appendix.  As in Eq.\eqref{eq:vars_optr} all kinematic variables
linked to the top tag will be evaluated with the fat jet size
$R_\text{opt}$.  The details of implementation of the $N$-subjettiness
variables is discussed in the Appendix.

%%%%%%%%%%%%%%%%%%%%%%%%%%%%%%%%%%%%%%%%%%%%%%%%%%%%%%%%%%%%%%%%%%%%%%
\subsection*{Qjets}

The main limitation even of the deterministic multivariate
\textsc{HEPTopTagger} is the aim to identify a unique set of subjets
from the top decay as part of the tagging procedure, which allows us
to reconstruct the 4-momentum of the tagger top and for example
compare it to Monte Carlo truth. If the kinematic selection identifies
a wrong set of subjets as the best candidates for the top decay
products, an actual top decay can easily fail the tagging
procedure. To avoid this loss in signal efficiency we can allow
for more than one set of candidate subjets to be tested. One approach
that not only covers several candidates of subjet combinations, but
which even allows for a statistical analysis of many such assignments
is \textsc{Qjets}~\cite{qjets}.\bigskip

During the clustering of the fat jet the standard recombination
algorithms combine the closest set of pre-jets according to a given
measure. For the C/A algorithm this measure is the
geometric separation $d_{ij} = \Delta R_{ij}^2$ of the pre-jets $i$
and $j$. \textsc{Qjets} generalizes this deterministic choice to a
likelihood measure.  For each pair of pre-jets ($i,j$) it computes the weight
\begin{alignat}{5}
\omega_{ij}^{(\alpha)}
&= \exp 
  \left( - \alpha \; \frac{d_{ij} - d_{ij}^\text{min}}{d_{ij}^\text{min}} 
  \right) \; ,
\label{eq:qjets1}
\end{alignat}
and then chooses the two pre-jets to cluster according to a random
number trailing the weights $\omega_{ij}^{(\alpha)}$.  For this study
we choose $\alpha = 0.1$, to balance the convergence of the algorithm
with our aim of generating alternative subjet assignments for the top
tagger. The standard jet algorithm corresponds to the limit $\alpha
\to \infty$. The global weight for a clustering history is defined as
\begin{alignat}{5}
\Omega^{(\alpha)}
&= \prod_\text{mergings} \omega_{ij}^{(\alpha)}
 = \left[ \prod_\text{mergings} \exp  \left( - \frac{d_{ij} - d_{ij}^\text{min}}{d_{ij}^\text{min}} 
  \right) \right]^\alpha 
\stackrel{\text{consistent}}{\longrightarrow} 1 \; .
\label{eq:qjets2}
\end{alignat}
The universal limiting case $\Omega^{(\alpha)} \to 1$ for a perfect
clustering history indicates that in searching for the largest global
weight $\Omega$ the choice of $\alpha$ should not make a major
difference. The \textsc{Qjets} clustering procedure can be repeated
many times, where in this study we typically rely on 100 clustering
histories.  They can be ranked by their global weights
$\Omega^{(\alpha)}$ instead of the independent local weights used by a
deterministic jet algorithm. For each history we apply the
unclustering and top tagging algorithm.  As long as the deterministic
jet algorithm picks a reasonable merging history for a signal event
we expect the outcome of the deterministic tagger and the tagger
acting on the clustering history with the highest global weight to be
close.\bigskip

The first advantage of \textsc{Qjets} appears when during an early
clustering step the deterministic measure $d_{ij}$ identifies the
wrong merging in the sense that the remaining history cannot be
described well by QCD. This deterministic history will by definition
receive the maximum global weight $\Omega^{(\alpha)} = 1$. However, an
alternative history in better agreement with QCD could reach a
similarly large global weight.  Because \textsc{Qjets} provides many
alternative clustering histories, we can search for a set of top tags
with comparably large global weights. For example, we can use the two
positively tagged \textsc{Qjets} histories with the highest global
weight in the multivariate analysis. This way, a possibly misleading
deterministic result is corrected.  This should improve the
performance in particular when we enforce high signal efficiencies,
where the tagger becomes most vulnerable to a wrong clustering input.
It turns out that already this simple modification gives a sizeable
improvement in the signal efficiency.

The second improvement to the usual top tagger is based on
\textsc{HEPTopTagger} output for the full set of 100 clustering
histories.  First, we include the fraction of positive top tags based
on the default \textsc{HEPTopTagger} settings among all 100
\textsc{Qjets} histories, $\eps_\text{Qjets}$, as introduced in the
Appendix. Next, we extract statistical information from distributions
of the \textsc{Qjets} histories, like for example the reconstructed
top mass $m_\text{rec}$. This distribution is defined for
$\eps_\text{Qjets} \times 100$ histories.  Signal events will strongly
peak around the top mass with a possible secondary peak around the
$W$-mass. QCD background events will instead show a smooth
decrease. The two most relevant observables in the $m_\text{rec}$
distribution are the mean and the variance of this reconstructed top mass
distribution with 100 entries, symbolically denoted as
$\{ m_\text{rec}^\text{Qjets} \}$.\bigskip

Our multivariate analysis we base on the second approach. We start
with the top-tagged \textsc{Qjets} history with the highest global weight and run
the tagging algorithm of this history only. In addition, we include
the statistical information of the $m_\text{rec}$ distribution of the
subset of the 100 \textsc{Qjets} histories which defines a top
candidate.  The complete list of observables including the
\textsc{Qjets} information now reads
\begin{equation} 
\{ \; m_{tt}, m_\text{ff}, p_{T,t_1}, p_{T,t_2}, 
      p_{T,\text{f}_1}, p_{T,\text{f}_2}, 
      m_\text{rec}^\text{min}, m_\text{rec}^\text{max}, f_\text{rec}^\text{max},
      R_\text{opt} - R_\text{opt}^\text{(calc)}, \{ \tau_N \},
      \eps_\text{Qjets}^\text{min}, \{m_\text{rec}^\text{Qjets}\} \; \} 
\qquad \text{(\textsc{Qjets}),}
\label{eq:vars_qjets}
\end{equation}
where $\{ \tau_N \}$ represents the appropriate set of filtered and
unfiltered $N$-subjettiness variables (for example $N=1,2,3$ for each
of the two tops).  For the two tags in the $Z'$ analysis we choose the
smaller $\eps_\text{Qjets}$ value of the two.  All variables from the
tagger are evaluated for the optimized $R$ size and the clustering
history with the largest global weight.\bigskip

In Fig.~\ref{fig:roc_new} we show the effect of the \textsc{Qjets}
histories in addition to the other improvements. A key difference
between the previous discussion and the \textsc{Qjets} approach is
that we now need to include some kind of detector resolution, to limit
\textsc{Qjets} to a manageable number of significantly different
merging histories. For that reason we divide the calorimeter into
$\eta\times\phi$ cells of size $0.1 \times 0.1$ and pre-cluster the
entire set of calorimeter entries before applying any jet algorithm.
Because this detector resolution effect is not included for the
previous results, the \textsc{Qjets} ROC curve does not consistently
exceed the $N$-subjettiness curve without \textsc{Qjets}. On the
other hand, we still observe the expected improvement towards large
signal efficiencies. The moderate drop at small signal efficiencies
gives us confidence that a full detector simulation will not lead to
significant degradation of our results.

%%%%%%%%%%%%%%%%%%%%%%%%%%%%%%%%%%%%%%%%%%%%%%%%%%%%%%%%%%%%%%%%%%%%%%
\section{Full event information}
\label{sec:comparison}

Going back to the discussion in Sec.~\ref{sec:resonance} the 
remaining question is how the new \textsc{HEPTopTagger2} performance
compares to other approaches designed for the upcoming LHC run. The
benchmark for such a comparison is event deconstruction, or more
specifically the projections for a $Z'$ resonance
search~\cite{event_deco}. As mentioned in our discussion of jet
radiation in Sec.\ref{sec:resonance} the borders between the hard
process or the $Z'$ decay on the one side and QCD jet
radiation and its sensitivity to the signal and background color
structure on the other side are washed out when we include for example
filtered subjets or $N$-subjettiness information. We therefore start
with a brief discussion of the additional information from jets in the
entire event and then move on to the comparison with the leading
benchmark in proposed $Z'$ analyses.

%%%%%%%%%%%%%%%%%%%%%%%%%%%%%%%%%%%%%%%%%%%%%%%%%%%%%%%%%%%%%%%%%%%%%%
\subsection*{Additional jets}

To determine to what degree the jet structure of purely hadronic $Z'
\to t\bar{t}$ events helps the extraction of the signal from the
$t\bar{t}$ and QCD jets background we first study the number and
kinematic distribution of small C/A jets with $R=0.2$ and $p_{T,j} >
10~\gev$ in addition to the fat jets fulfilling
Eq.\eqref{eq:fat_cuts}. We choose these very small jets in order to
test information which might be available from so-called microjets
in shower deconstruction. Our discussion should not
be applied to an LHC analysis one-to-one and is instead aimed at
capturing as much information as possible. Without any major cuts, the
number of jets will consist of three decay jets per top quark, FSR
jets, and ISR jets. For an inclusive event sample, we should be able
to tell apart the different processes from the number of jets and the
kinematics of the individual jets~\cite{scaling}.

\begin{figure}[t]
  \includegraphics[width=0.32\textwidth]{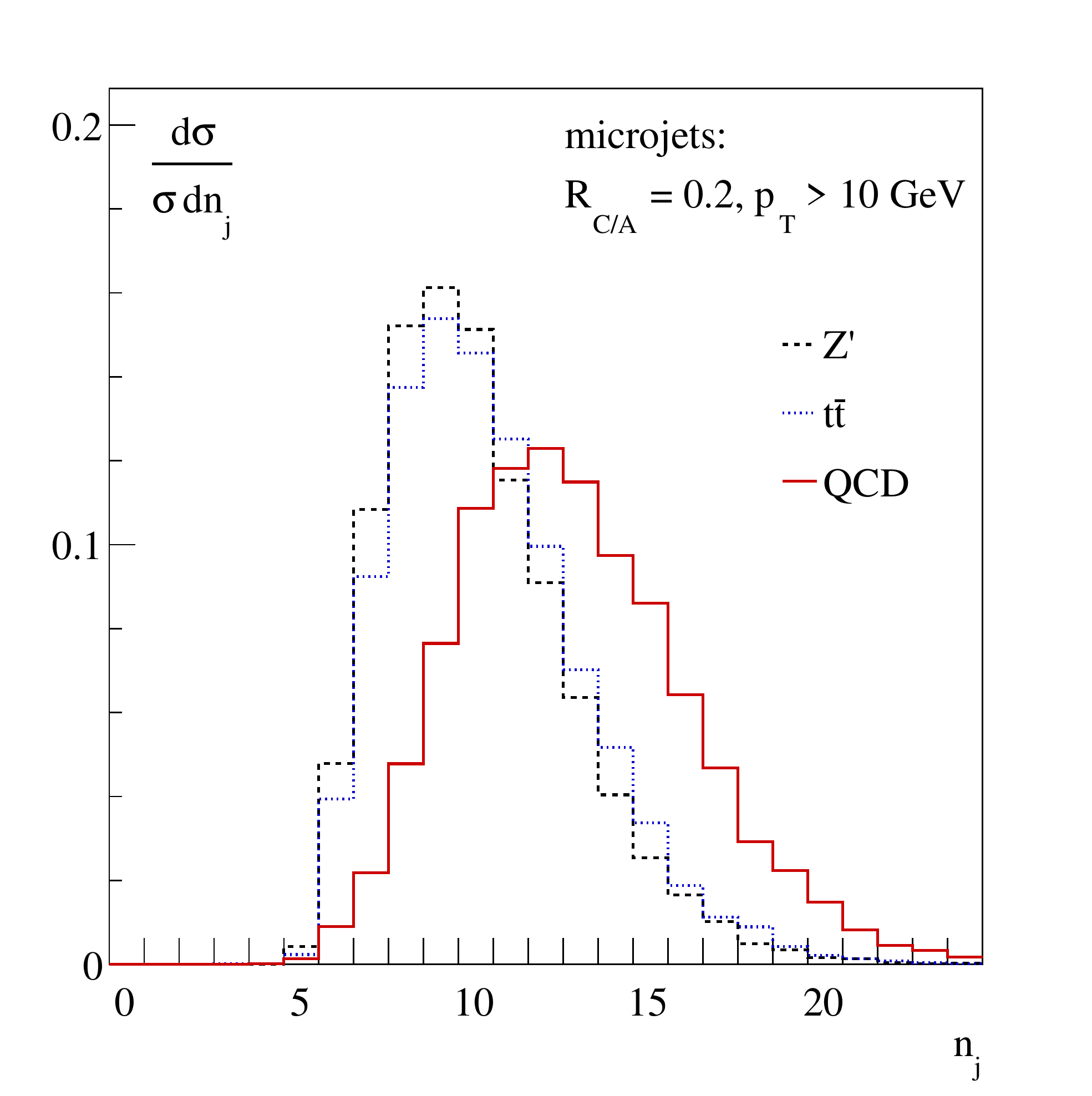}
  \includegraphics[width=0.32\textwidth]{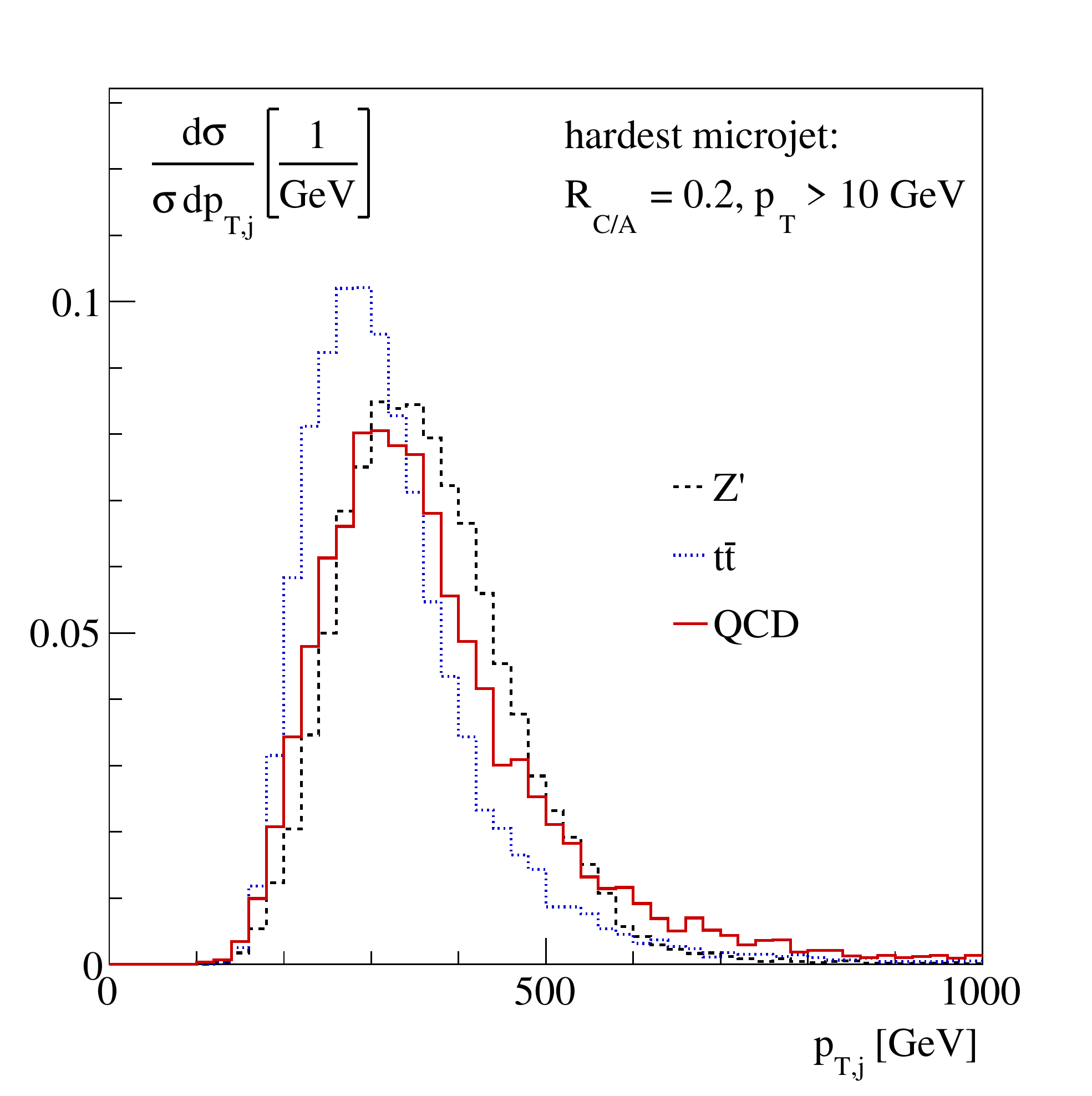}
  \phantom{\includegraphics[width=0.32\textwidth]{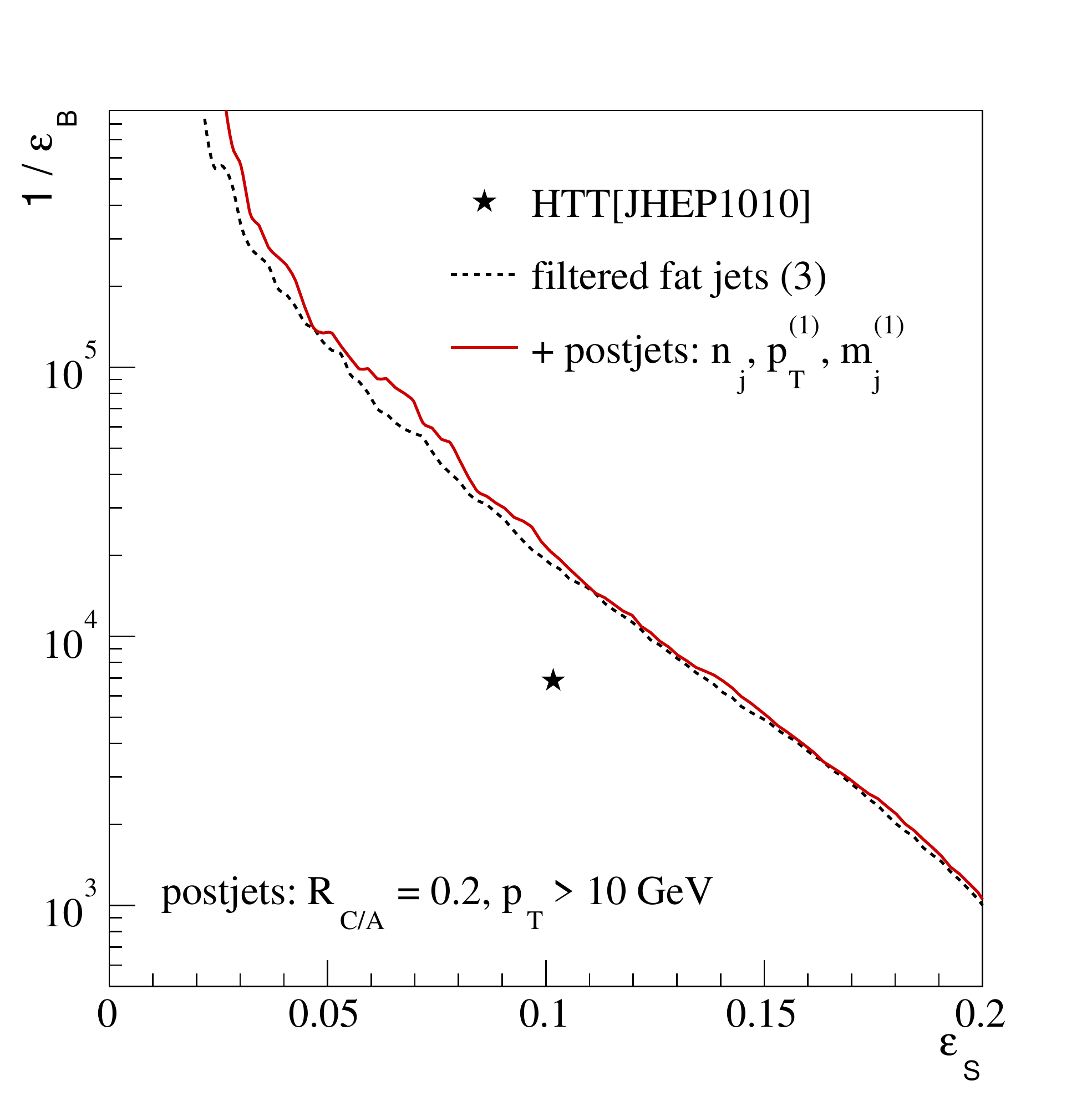}} \\
  \includegraphics[width=0.32\textwidth]{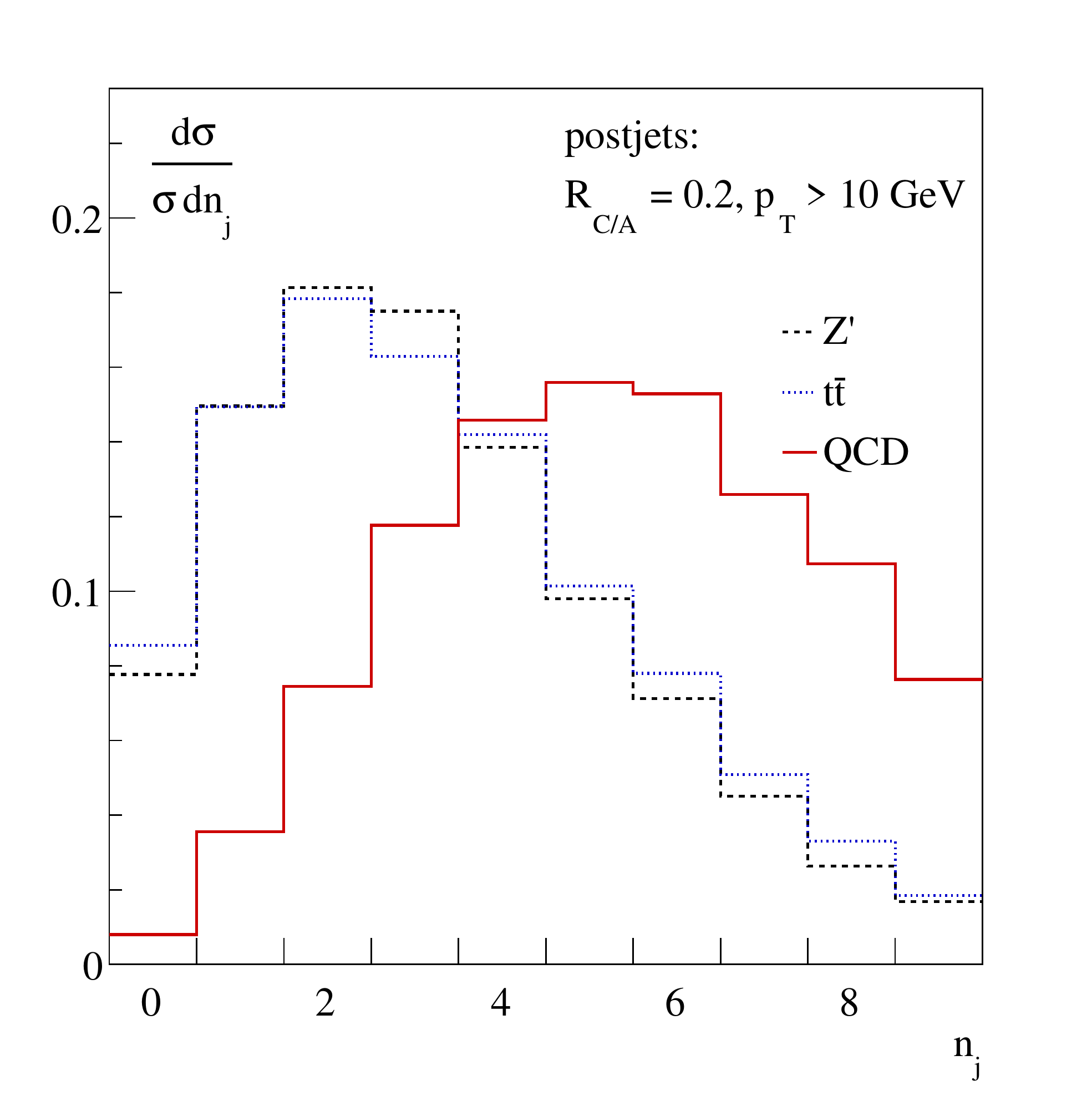}
  \includegraphics[width=0.32\textwidth]{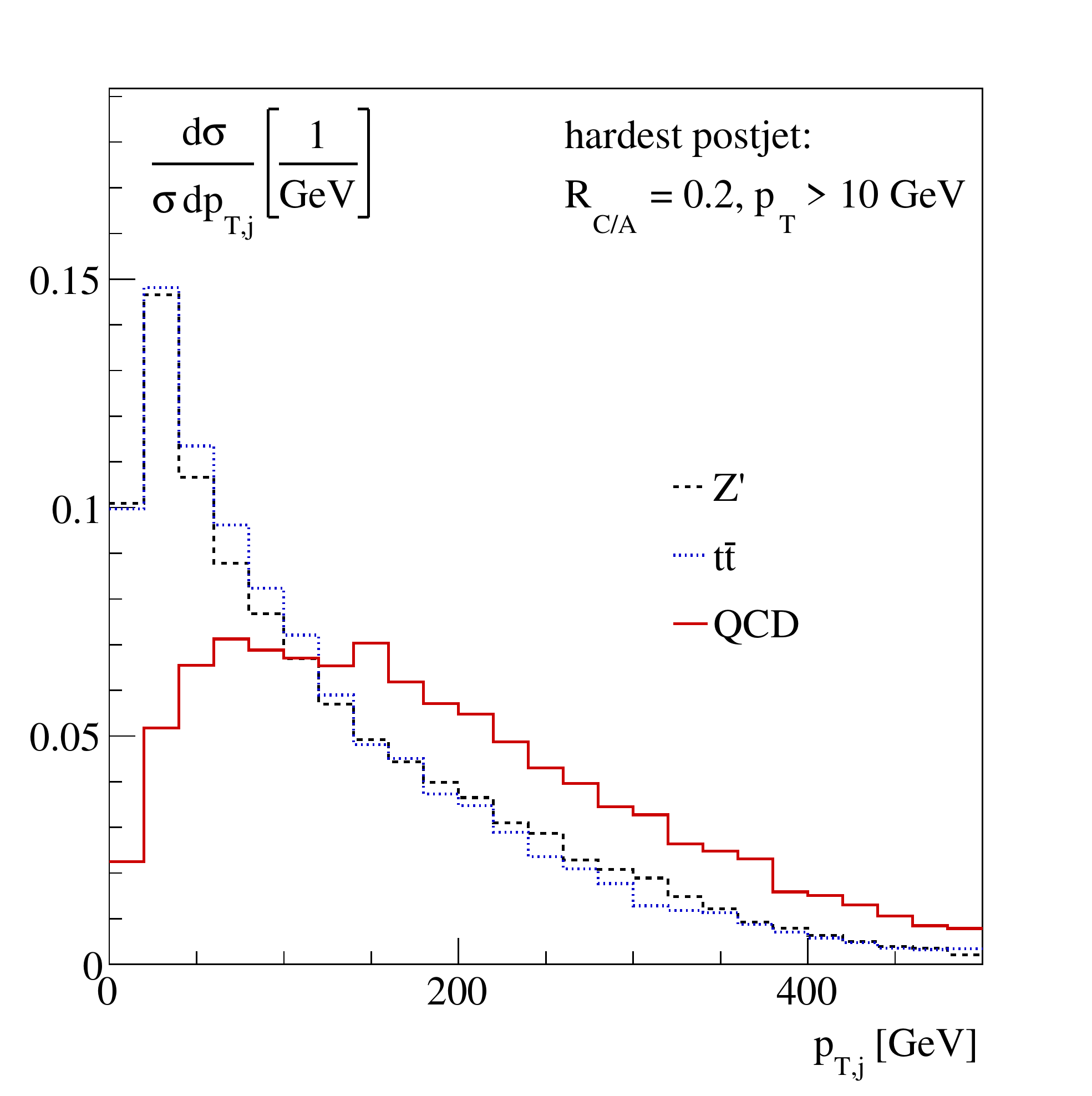}
  \includegraphics[width=0.32\textwidth]{roc_addjet} 
  \caption{Information on the hardest jet before top tagging (upper
    row) and the hardest jet left over after top tagging (lower row).
    For the jets defined with $R=0.2$ and $p_T>10~\gev$ we show the
    number of jets, the hardest jet's transverse momentum, and its
    mass in $Z'$ candidate events (left to right).}
  \label{fig:jets}
\end{figure}

After a first level of cuts we see in Fig.~\ref{fig:jets} that the
$Z'$ signal and the $t\bar{t}$ background both peak at 10 microjets,
\eg four jets from ISR and FSR combined. For the background the number
is slightly larger, because we generate the scale of the hard process
also through a large number of jets. We also see that the transverse
momentum of the hardest jet is slightly larger for the signal. We
could include these jet patterns in a multivariate analysis, but at
this stage this information would be very heavily correlated with the
variables from the top tagger.\bigskip

In a second step we focus on the jet activity which does not
contribute to the top tagging.  Inside the fat jets we know that the
top tagger includes information based on subjets with typically
$R=0.3$ and $p_T \gtrsim 20~\gev$ after filtering.  After two tags we
then remove all calorimeter data associated with the filtered triplet
of either of the top candidates and re-cluster the remnants into
microjets with $R=0.2$ and $p_{T,j} > 10~\gev$.  In the lower panels
of Fig.~\ref{fig:jets} we see how after removing the signal decay jets
the remaining number of jets peaks around two ISR or FSR jets. For the
QCD background this number is higher, because it takes a larger number
of equally distributed jets in the detector to fake a boosted massive
top inside each fat jet. The transverse momentum of the hardest of the
remaining QCD jets also peaks at very small values for the signal and
the $t\bar{t}$ background, as one would expect for example for a small
number of ISR jets. The bulk of the hardest QCD jets per event shows
transverse momenta around $p_{T,j} = 50-200~\gev$, still small
compared to the hard scale imprinted on the multi-jet background
through the kinematic selection of Eq.\eqref{eq:fat_cuts}.  We should
be able to use this additional information for our BDT analysis, to
improve the signal extraction. In the right panel of
Fig.~\ref{fig:jets} we see the corresponding ROC curve. It turns out
that almost all of the information available through the extra jet
radiation is already included in our combined analysis of top tags and
subjet kinematics.

Based on this piece of information we assume that additional jet
information inside and outside the fat jets hardly changes the stable
results of the updated top tagger, so we can compare the new
\textsc{HEPTopTagger2} to other multivariate methods.

%---------------------------------------------------------------------
%---------------------------------------------------------------------

%%%%%%%%%%%%%%%%%%%%%%%%%%%%%%%%%%%%%%%%%%%%%%%%%%%%%%%%%%%%%%%%%%%%%%
\subsection*{Comparison with other approaches}

The most promising projections for boosted top identification and
specifically searches for $t\bar{t}$ resonances during the upcoming
LHC runs are available for shower deconstruction~\cite{shower_deco} or
event deconstruction~\cite{event_deco}. This method is based on a
construction of likelihoods representing possible shower histories for
a jet or a fat jet. The underlying objects are so-called
C/A~\cite{ca_algo,fastjet} microjets with $R=0.2$ and $p_T >
10~\gev$~\cite{event_deco}. They are slightly softer and smaller than
the subjets in a typical top tagger, but we have seen that the
additional information from those jets should not make a big
difference. Unlike general template methods, shower deconstruction
relies on the soft and/or collinear approximation of QCD to compute
the likelihood of a given shower history in terms of splitting
probabilities and Sudakov factors (non-splitting
probabilities). Based on the possible shower histories the likelihood
ratio of a fat jet coming from a boosted top quark or from the QCD jet
background acts as a measure for the top tag.  One problem with shower
deconstruction, like any probabilistic approach, is that we cannot
separate the identification and the reconstruction of the boosted top
quark. This means we cannot for example show the quality of the
reconstructed 4-momentum compared to Monte Carlo truth.\bigskip

The $Z'$ analysis using event deconstruction starts with two fat jets
of size $R=1.5$ and the acceptance cuts given in
Eq.\eqref{eq:fat_cuts}. The number of microjets is limited to 9 per
fat jet. In addition to the likelihood separating the top or QCD
origin of each of the two fat jets, the event likelihood measure now
also includes a likelihood describing the resonant or non-resonant
production of the pair of fat jets given their 4-momenta. At the level
of the hard process this part is not very different from the
established matrix element method~\cite{me_method} and largely
replaces an analysis of the $m_{tt}$ and $p_{T,t}$ distributions
defining the multivariate analysis of Eq.\eqref{eq:vars_kin}. In
Ref.~\cite{event_deco} the observable width of the $m_{tt}$ resonance
is assumed to range around $65~\gev$, an assumption we follow. In our
analysis the precise resolution for example after detector effects
only plays a secondary role, because the resolution of the
\textsc{HEPTopTagger2} is limited to $145~\gev$, as shown in
Tab.\ref{tab:fit}.

%---------------------------------------------------------------------
\begin{figure}[t]
  \begin{center}
  \includegraphics[width=0.4\textwidth]{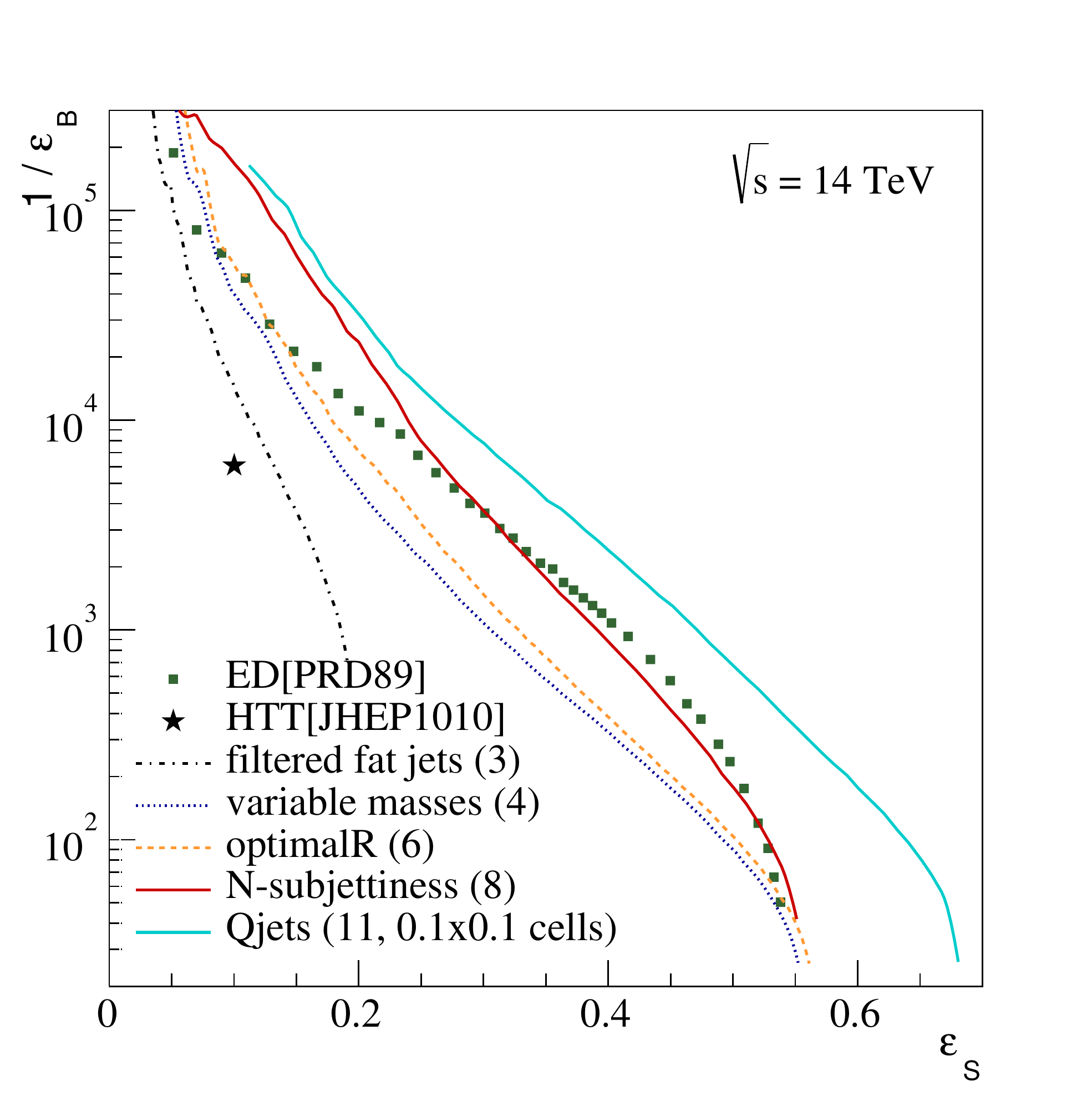}
  \end{center}
  \caption{Comparison of the multivariate \textsc{HEPTopTagger2}
    analysis presented in this paper with the event deconstruction
    approach of Ref.~\cite{event_deco}.  All \textsc{HEPTopTagger2}
    curves correspond to Fig.~\ref{fig:roc_new}, but now with a
    collider energy of 14~TeV instead of 13~TeV, This comparison in
    the absence of an experimental validation should be taken as first
    estimate.}
\label{fig:comparison}
\end{figure}
%---------------------------------------------------------------------

In Fig.~\ref{fig:comparison} we show the performance of the analysis
developed in this paper with the recent benchmark of event
deconstruction. One difference to the \textsc{HEPTopTagger} results
shown in Fig.~\ref{fig:roc_new} is that we now show $Z'$ efficiencies
up to 68\%, confirming that \textsc{Qjets} indeed gives us a major
improvement for very large signal efficiencies.  Another difference is
that for a direct comparison we now assume a collider energy of
14~TeV.  Both, event deconstruction and the new \textsc{HEPTopTagger}
show a comparable performance for the upcoming run.  The final answer
on both methods will only be given by experimental studies including
data.

%%%%%%%%%%%%%%%%%%%%%%%%%%%%%%%%%%%%%%%%%%%%%%%%%%%%%%%%%%%%%%%%%%%%%%
\section{Conclusion}
\label{sec:conclusion}

We demonstrated how the updated \textsc{HEPTopTagger2} performs in
searches for $Z'$ bosons or other heavy resonances decaying to top
pairs in the upcoming LHC run. Based on the original
\textsc{HEPTopTagger}~\cite{heptop2} we modify the tagging algorithm
and add several additional kinematic variables to a multivariate
analysis:
\begin{itemize}
\item[--] fat jet kinematics to account for final-state radiation in
  resonance searches;
\item[--] algorithmically optimized size of the original fat jet combined
  with its prediction (optimalR mode);
\item[--] $N$-subjettiness probing the more general subjet structures
  inside the fat jet;
\item[--] \textsc{Qjets} with a global picture of the most likely
  clustering histories giving a top tag.
\end{itemize}
Each of these improvements can be added to the top tagging
individually. For the specific $Z'$ resonance search we altogether
achieve an increase of the background rejection by a factor of $30$
for a constant $Z'$-signal efficiency of $10\%$. Compared to the
original tagger~\cite{heptop2} the background sculpting in the
invariant mass of the top pair is significantly
reduced~\cite{heptop4}.  These updated results are at least
competitive with the leading estimates for other tagging
methods.\bigskip

Because the multivariate $Z'$ analysis includes several layers of
improvement, not necessarily linked to the actual top tagging, we also
show in the Appendix the corresponding improvements for top tagging in
$t\bar{t}$ events.  There, we test the updated tagger for moderate
($p_{T,t}>200$~GeV) and sizeable ($p_{T,t}>600$~GeV) boost and find a
significant improvement in particular for larger boost. The limiting
factor for moderate boost still is capturing all three top decay jets
inside a fat jet, which has to be targeted by a dedicated low-$p_T$
mode~\cite{heptop3}. The corresponding \textsc{HEPTopTagger2}
described in the Appendix will be made publicly
available~\cite{heptop2,fastjet_contrib}. In particular for
\textsc{Qjets} there exist different modes which need to be tested on
data.

Comparing the improvement of the $Z'$ analysis with that in the
individual top tags shows that the benefits for the full $Z'$ case are
significantly larger than those just from the top tags. A lesson from
this is that it is useful to consider the optimization of top tagging,
not only in its own right, but also in the context of full search
analyses.\bigskip

%%%%%%%%%%%%%%%%%%%%%%%%%%%%%%%%%%%%%%%%%%%%%%%%%%%%%%%%%%%%%%%%%%%%%%
 \begin{center}
 {\bf Acknowledgments}
 \end{center}
 T.S. acknowledges support by the IMPRS for Precision Tests of
 Fundamental Symmetries. 
 GPS acknowledges partial support from ERC advanced grant Higgs@LHC.

%%%%%%%%%%%%%%%%%%%%%%%%%%%%%%%%%%%%%%%%%%%%%%%%%%%%%%%%%%%%%%%%%%%%%%
\clearpage
\appendix

%%%%%%%%%%%%%%%%%%%%%%%%%%%%%%%%%%%%%%%%%%%%%%%%%%%%%%%%%%%%%%%%%%%%%%
\section*{Appendix: HEPTopTagger2}
\label{app:algo}

In the past it has proven useful to publish details about the
\textsc{HEPTopTagger} algorithm. We describe the new structure
reflecting all changes in Refs.\cite{heptop2,heptop3,heptop4} in this
Appendix. Because the main body of the paper is focused on the
performance in resonance searches we then present benchmark
results based on purely hadronic $t\bar{t}$ events in the Standard
Model. They can be directly translated for example into
semi-leptonically decaying $t\bar{t}$ pairs. Finally, the enhanced
capabilities of the \textsc{HEPTopTagger2} have lead to enough of a
complexity of the actual code that we briefly describe the run modes,
the input parameters, and the available output information from the
tagger.

%%%%%%%%%%%%%%%%%%%%%%%%%%%%%%%%%%%%%%%%%%%%%%%%%%%%%%%%%%%%%%%%%%%%%%
\subsection*{Algorithm}

The basic \textsc{HEPTopTagger2} algorithm largely follows the
original algorithm described in Ref.~\cite{heptop2}, but is based on
\textsc{FastJet3}~\cite{fastjet} and includes a number of new
features:
\begin{enumerate}
\item define a C/A fat jet with $R_\text{fat}=1.8$ and determine the
  splitting history through the default clustering.
\item identify all hard subjets using a mass drop criterion: undo the
  last clustering of the jet $j$, into two subjets $j_1,j_2$ with
  $m_{j_1} > m_{j_2}$; require $m_{j_1} < f_\text{drop}~m_j$ with
  $f_\text{drop} = 0.8$ to keep both; otherwise, keep only $j_1$;
  further decompose or add each subjet $j_i$ to the list of relevant
  substructures. A global soft cutoff $m_{j_i} > m_\text{min} =
  30~\gev$ can be adjusted\footnote{We have checked that replacing the
    mass drop criterion with a soft drop criterion~\cite{soft_drop}
    does not improve the performance of the tagger noticeably.}.
\item iterate through all triplets of three hard subjets: filter them
  with resolution $R_\text{filt}=\min(0.3,\Delta R_{jk}/2)$; use the
  $N_\text{filt} = 5$ hardest filtered constituents and calculate
  their combined jet mass; re-cluster these five subjets into three
  assumed top decay jets; reject all triplets outside $m_{123} \equiv
  m_\text{rec} \in [150,200]~\gev$; keep the event if at least one
  such triplet exists. For the multivariate analysis this window is
  opened to $m_\text{rec} < 1~\tev$, which allows us to use
  $m_\text{rec}$ as a kinematic output of the tagger.

  This set of re-clustering and filtering steps by default uses
  the C/A jet algorithm~\cite{ca_algo}. However, to guarantee infrared
  safety and enhance the performance at large boosts~\cite{heptop3} it
  can be switched to $k_T$ jets~\cite{kt_algo}.

\item order the three subjets $j_1, j_2, j_3$
  by $p_T$; if the masses
  $(m_{12}, m_{13},m_{23})$ satisfy one of the following three
  criteria, accept them as a top candidate:
\begin{alignat}{5}
&0.2 <\arctan \frac{m_{13}}{m_{12}} < 1.3
\qquad \text{and} \quad
R_\text{min}< \frac{m_{23}}{m_{123}} < R_\text{max}
\notag \\
&R_\text{min}^2 \left(1+\left(\frac{m_{13}}{m_{12}}\right)^2 \right) 
< 1-\left(\frac{m_{23}}{m_{123}} \right)^2
< R_\text{max}^2 \left(1+\left(\frac{m_{13}}{m_{12}}\right)^2 \right)  	
\quad \text{and} \quad 
\frac{m_{23}}{m_{123}} > 0.35
\notag \\
&R_\text{min}^2\left(1+\left(\frac{m_{12}}{m_{13}}\right)^2 \right) 
< 1-\left(\frac{m_{23}}{m_{123}} \right)^2
< R_\text{max}^2\left(1+\left(\frac{m_{12}}{m_{13}}\right)^2 \right)  	
\quad \text{and} \quad 
\frac{m_{23}}{m_{123}}> 0.35
\label{eq:app_select}
\end{alignat} 
  where $R_\text{min,max}= (1 \mp f_W) m_W/m_t$ defines the parameter
  $f_W$, by default set to $f_W = 0.15$. The soft cutoff $m_{23} >
  0.35~m_{123}$ as well as the limits $[0.2,1.3]$ in the first line
  can be adjusted.  All kinematic cuts are listed in
  Tab.~\ref{tab:htt_params} and can be adapted in a multivariate
  approach. In the multivariate case we open the $W$-mass window
    to $f_W = 0.3$.  The ratio of the $W$-mass to the top mass can
    then be used as a kinematic output defined as
\begin{equation}
 f_\text{rec} = 
  \min_{ij} \left| \frac{\; \dfrac{m_{ij}}{m_\text{123}} \;}{\dfrac{m_W}{m_t}} - 1 \right| 
\label{eq:f_rec}
\end{equation}

\item of all triplets passing the above criteria in a given fat jet
  choose the one with $m_{123} \equiv m_\text{rec}$ closest to
  $m_t$. This selection has shown to be the most efficient, and
  applying it after all kinematic cuts minimizes the background
  sculpting.  The $m_\text{rec}$ and $f_\text{rec}$ values supplied to
  the multivariate analysis are those corresponding to this triplet.

\item for consistency, require the reconstructed $p_{T,t}$ to exceed
  $200~\gev$.
\item in the low-$p_T$ mode~\cite{heptop3} reduce this threshold to
  $p_{T,t} > 150~\gev$; compute the Fox--Wolfram moments~\cite{fwm}
\begin{alignat}{2}
H^x_\ell &= \sum_{i,j=1}^N \; W_{ij}^x \; P_\ell(\cos \Omega_{ij}) \notag \\
\text{with}& \quad W_{ij}^T = \frac{p_{T i}\,p_{T j}}{\left(\sum p_{T i}\right)^2}  
\qqquad \text{and} \qqquad W_{ij}^U = \frac{1}{N^2} \; .
\label{eq:fwm}
\end{alignat}
  of the subjets relative to each other and relative to the
  reconstructed top momentum.  This mode is not part of the usual
  tagger and relies on external \textsc{GSL} libraries~\cite{gsl} for Legendre
  polynomials.
\item in the optimalR mode repeat steps~1 to~3 with a decreasing fat
  jet radius in steps of $\Delta R = 0.1$; based on the condition
  $m_\text{rec}^{(1.8)} - m_\text{rec} > 0.2 m_\text{rec}^{(1.8)}$
  determine the minimum radius $R_\text{opt} > 0.5$; follow steps~4
  to~6 with this modified fat jet. We also parametrize the
  expected value for $R_\text{opt}$ in terms of
  $p_{T,\text{f}}$ based on the numerical simulation of the top decay kinematics 
  illustrated in Fig.~\ref{fig:r_opt_fit}
\begin{equation}
R_\text{opt}^\text{(calc)} = \frac{327}{p_{T,\text{f}}} \; .
\label{eq:ropt_fit}
\end{equation} 

%---------------------------------------------------------------------
\begin{figure}[b!]
  \includegraphics[width=0.4\textwidth]{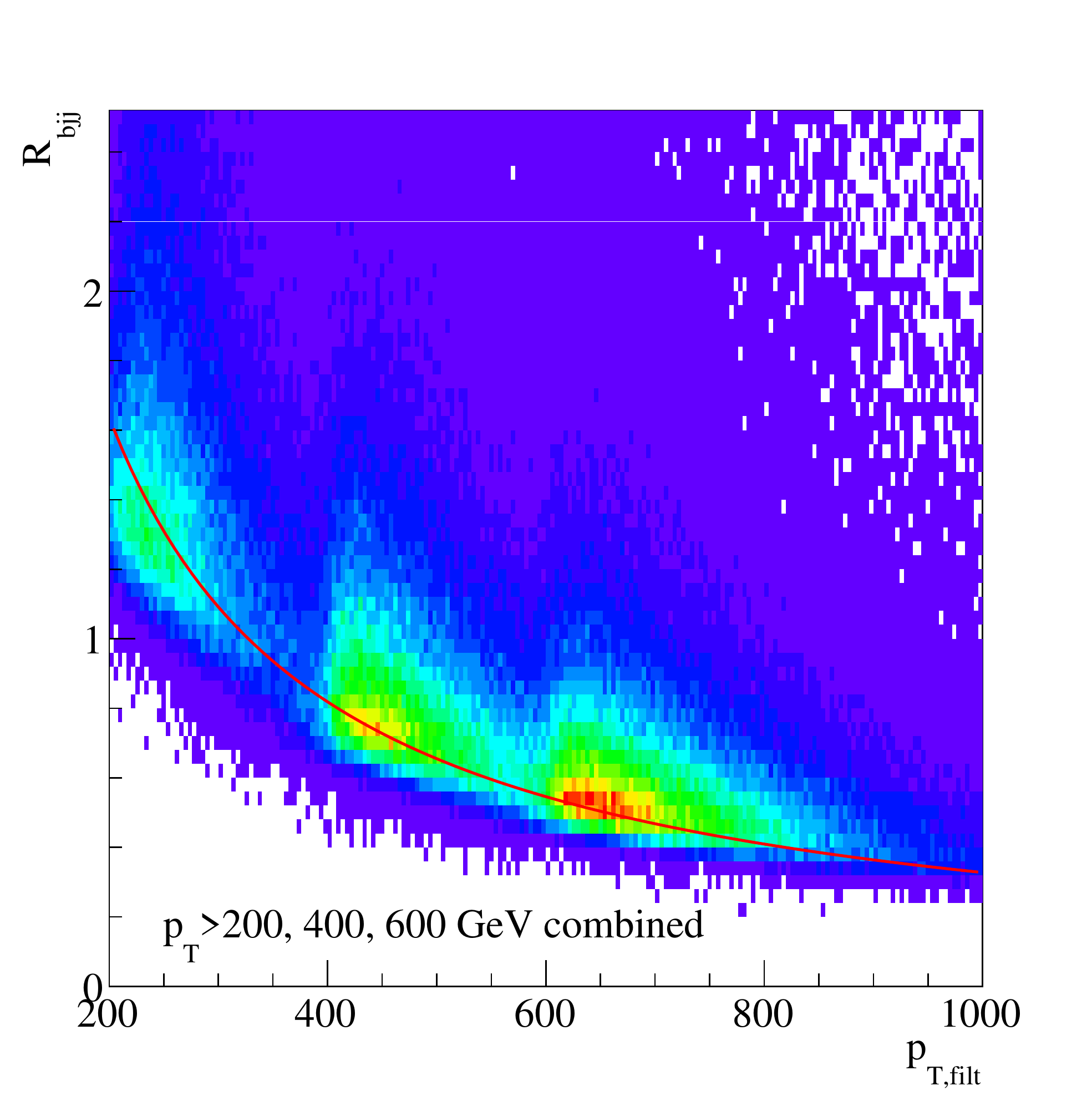}
  \caption{$R_\text{opt}^\text{(calc)}$ fit based on Standard Model
    $t\bar{t}$ samples with $p_{T,t} > 200, 400, 600$~GeV for the
    parton level distance of decay products $R_{bjj}$. The fat jets
    are filtered with $R=0.2$, $N=10$. The functional form of the fit
    curve is given in Eq.\eqref{eq:ropt_fit}.}
  \label{fig:r_opt_fit}
\end{figure}
%---------------------------------------------------------------------

\item in the $N$-subjettiness mode~\cite{heptop4} compute the
  $\tau_j$~\cite{nsubjettiness} as defined in Eq.\eqref{eq:tau_N} from
  the filtered and unfiltered subjets, as described below. Again, this
  mode is not part of our tagger code and relies on the 
  \textsc{FastJet Contrib}~\cite{fastjet,fastjet_contrib} add-on for
  $N$--subjettiness~\cite{nsubjettiness}.
\item in the \textsc{Qjets} mode replace the deterministic output of
  step~1 by a set of possible histories defined in
  Eq.\eqref{eq:qjets2}; run the tagger for each of them, giving a
  set of clustering histories with global weights $\Omega$, and a
  positive or negative tagging result.
\end{enumerate}
Following this description the low-$p_T$~(7) and $N$-subjettiness~(9)
modes simply add kinematic observables to the tagger output. These
observables can be included in a multivariate analysis or can be cut
on in the deterministic top tagging decision. The improvement in the
low-$p_T$ mode is illustrated in detail in Ref.~\cite{heptop4} while
the impact of $N$-subjettiness variables on the resonance search is
illustrated in Fig.~\ref{fig:roc_new}.

In contrast, the optimalR mode and the \textsc{Qjets} mode modify the
clustering histories~(1) underlying the mass drop
search~(2). Depending on the modified fat jet size or on the
\textsc{Qjets} weight they return a set of tagging outputs. For the
optimalR mode it is straightforward to choose the smallest reasonable
fat jet size $R_\text{opt}$ for the actual tagging. The \textsc{Qjets}
histories can be evaluated in a range of possible ways.

%%%%%%%%%%%%%%%%%%%%%%%%%%%%%%%%%%%%%%%%%%%%%%%%%%%%%%%%%%%%%%%%%%%%%%
\subsection*{Performance}

%---------------------------------------------------------------------
\begin{figure}[t!]
  \includegraphics[width=0.4\textwidth]{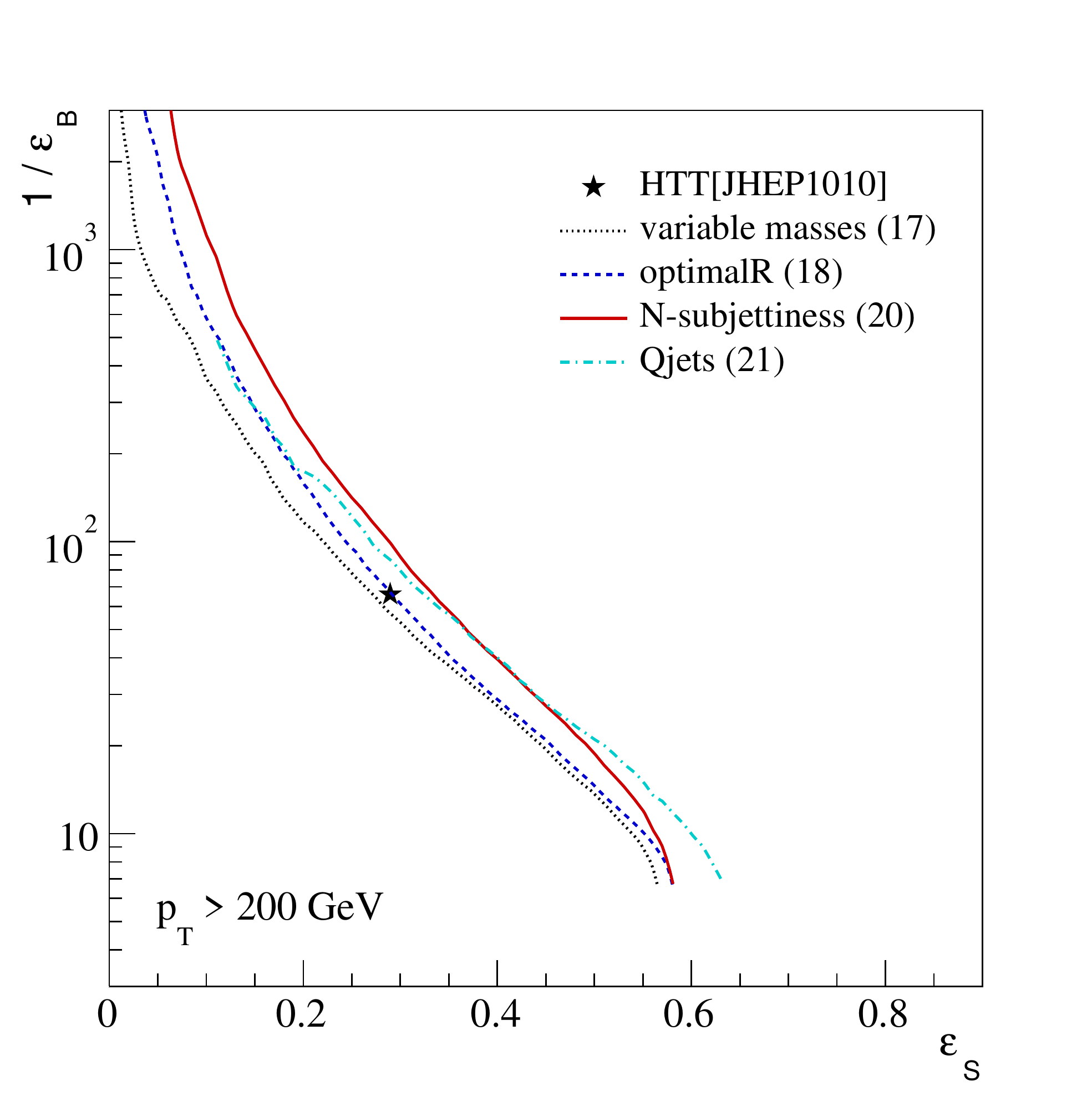}
  \hspace*{0.1\textwidth}
  \includegraphics[width=0.4\textwidth]{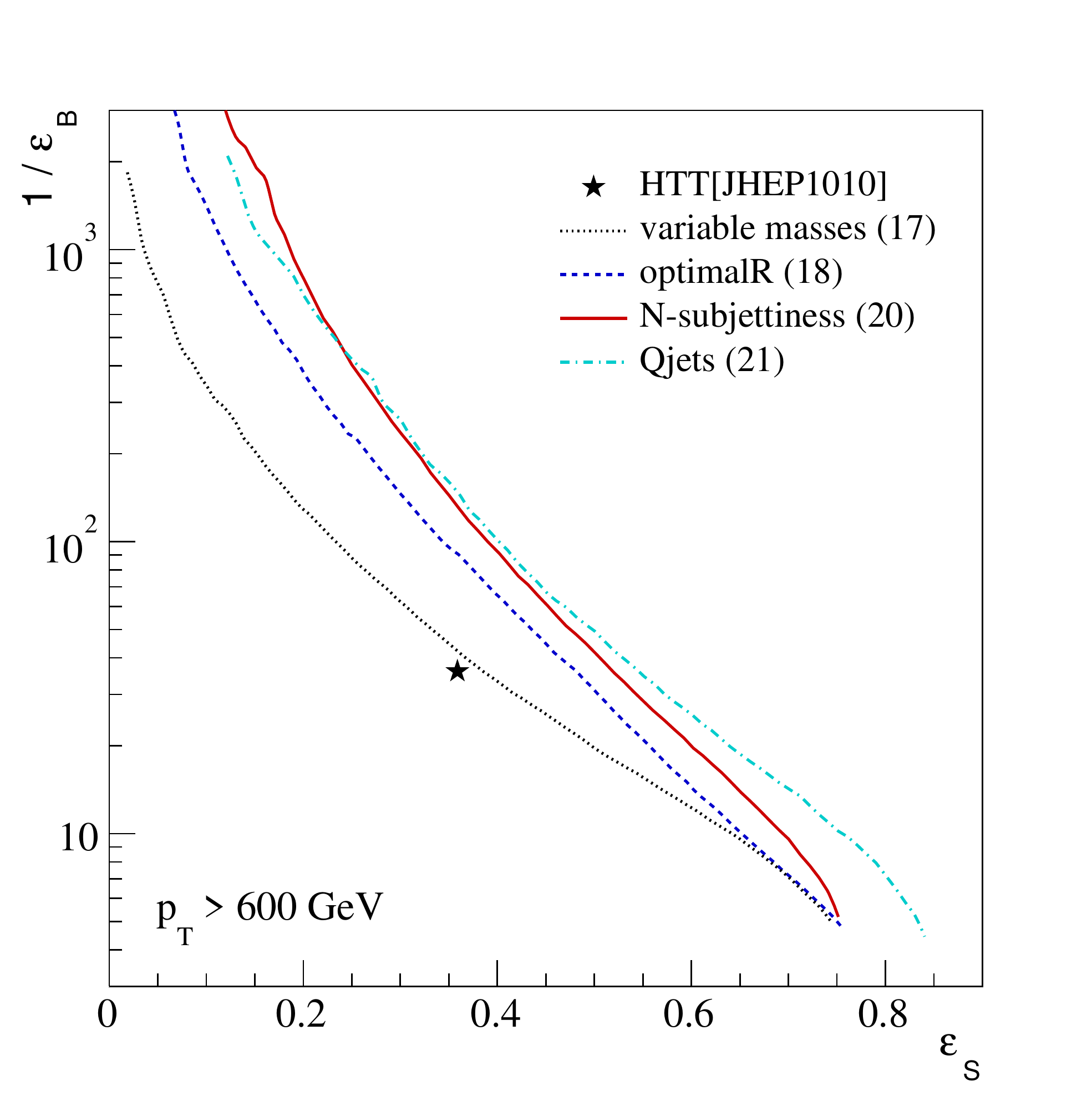} 
\caption{Performance of the \textsc{HEPTopTagger2} for $t\bar{t}$
  production in the Standard Model. We show the
  incremental improvements from the extended multivariate analyses for
  top quarks with $p_{T,t} > 200~\gev$ and $p_{T,t} > 600~\gev$.}
  \label{fig:performance}
\end{figure}
 %---------------------------------------------------------------------

The main body of this paper focuses on $t\bar{t}$ resonance searches
using the \textsc{HEPTopTagger} described above. While the combination
of tagged top kinematics and fat jet kinematics in
Sec.\ref{sec:resonance} does not directly translate into to a
universal top tagger, the multivariate aspects discussed in
Sec.~\ref{sec:optr}, namely optimalR, $N$-subjettiness, and
\textsc{Qjets} do.  Here, we show efficiencies for extracting
$t\bar{t}$ events from the QCD multi-jet background.  

Our analyses are based on fully hadronic $t\bar{t}$ signal and QCD
dijet background samples generated with
\textsc{Pythia8}~\cite{pythia}. For the general top tagger analysis 
in this Appendix we include underlying event in the
event generation and mimic the limited detector resolution by
clustering the hadronic activity into $\eta\times\phi$ cells of size
$0.1 \times 0.1$, similar to the \textsc{Qjets} results shown in Fig.~\ref{fig:roc_new}.  
Instead of the hard acceptance cuts in
Eq.\eqref{eq:fat_cuts} we now allow for softer fat jets. Two
multivariate BDT analyses focus on $t\bar{t}$ samples with 
\begin{equation}
p_{T,\text{fat}} > 200~\gev \qqquad 
|y_\text{fat}|<2.5 \qqquad 
p_{T,t} > 200, 600~\gev \; ,
\label{eq:fat_cuts_app}
\end{equation}
where the top momenta are evaluated on the Monte Carlo truth level.
We select events with fat C/A jets of radius $R_\text{fat}=1.8$ and
$|y_\text{fat}| < 2.5$ constructed with \textsc{FastJet}.\bigskip

Background efficiencies $\eps_B $ are defined as relative to the
number of those fat jets. For the signal efficiencies we require that
the fat jets can be matched to a parton level top quark within $\Delta
R < 0.8$.  Using the original version of the
\textsc{HEPTopTagger}~\cite{heptop2} we find for the $p_T > 600$~GeV
samples a signal efficiency of $\eps_S = 35.6\%$ and a mis-tagging
rate $\eps_B = 2.7\%$. The first change in the algorithm addresses the
signal efficiency and background sculpting. In the original algorithm
the triplet of subjets closest to the true top mass is selected and
only later the mass plane cuts are applied. Therefore, the tagger will
fail if this triplet does not pass the mass plane constraints and no
alternative triplet is analyzed. To eliminate this limitation, we
first apply the mass plane constraints and then pick the triple
closest to the top mass, as described above.

As in the main text we study further improvements of the tagger based
on ROC curves. To allow for such improvements we loosen the cuts of
the tagger to $m_\text{rec} < 1$~TeV and $f_W = 0.3$. The
initial set of BDT parameters in analogy to Eq.\eqref{eq:vars_window}
is
\begin{equation}
\{\ m_\text{rec},  f_\text{rec}\ \}\;
\qqquad \text{(variable masses).}
\end{equation} 
The large cone size of $R=1.8$ is not always appropriate, so the
optimalR mode optimizes the radius of each fat jet. Starting from the
initial cone size we stepwise reduce the size of the fat jet until the
criterion Eq.\eqref{eq:r_drop} indicates that we miss a top decay
jet. For the last stable $R$ size we run the usual tagging algorithm.
We can calculate the expected value $R_\text{opt}^\text{calc}$ for the
critical radius based on the transverse momentum of the filtered fat
jet.  For a fat jet originating from a top decay this prediction
should agree with the measured value, while for a background fat jet
the two are only strongly correlated when the entire subjet kinematics
is a perfect match to a top decay.  For the optimalR mode we set up a
BDT analysis with the observables
\begin{equation}
 \{ \ m_\text{rec},  f_\text{rec},
  R_\text{opt} - R_\text{opt}^{(\text{calc})} \ \}
\qqquad \text{(optimalR).}
\end{equation}
All tagging observables are evaluated for a fat jet with size
$R_\text{opt}$. In Fig.~\ref{fig:performance} we show the improvement
from the optimized size of the fat jet. Obviously, it is more
impressive for larger boost, while for $p_{T,t} > 200$~GeV the optimalR
mode hardly leads to a reduction in fat jet size.\bigskip

The $N$-subjettiness variables are best applied independently for fat
jets which would pass and would not pass the initial tagging
criterion. The optimalR working point
\begin{equation}
m_\text{rec} \in [150,200]~\gev \qqquad
f_\text{rec} < 0.175 \qqquad 
R_\text{opt} - R_\text{opt}^{(\text{calc})} < 0.3 \; ,
\label{eq:multi_wp}
\end{equation}
which corresponds to the signal efficiency $\eps_S = 0.22 (0.27)$
in Fig.~\ref{fig:performance}, defines these two categories.  Fat jets
passing Eq.\eqref{eq:multi_wp} can be assumed to include a complete
set of top decay products and are filtered with
$R_\text{filt}^{(1)} = 0.2$ and $N_\text{filter}^{(1)}=5$; fat jets
failing this criterion are instead filtered with
$R_\text{filt}^{(0)} = 0.3$ and $N_\text{filter}^{(0)}=3$. The
unfiltered $N$-subjettiness variables $\tau_i$ defined in
Eq.\eqref{eq:tau_N} and their filtered counter parts
$\tau^{(0)}_i,\tau_i^{(1)}$ are included up to $i \leq 3$. 
The reference axes are chosen as $k_T$-axes. We then set
up two independent BDTs with
\begin{alignat}{2}
&\{ \ m_\text{rec},  f_\text{rec},
R_\text{opt} - R_\text{opt}^{(\text{calc})}, m_\text{fat}^{(1)},  
\tau^{(1)}_3, \tau^{(1)}_3/ \tau^{(1)}_2,
\tau^{(1)}_2/ \tau^{(1)}_1, \tau_2,  \tau_3/ \tau_2, \tau_2/ \tau_1 \ \} 
\qquad \text{($N$-subjettiness, pass)} \notag \\
&\{ \ m_\text{rec},  f_\text{rec},
R_\text{opt} - R_\text{opt}^{(\text{calc})}, m_\text{fat}^{(0)}, 
\tau^{(0)}_3, \tau^{(0)}_3/ \tau^{(0)}_2,
\tau^{(0)}_2/ \tau^{(0)}_1, \tau_1,  \tau_3/ \tau_2, \tau_2/ \tau_1 \ \}
\qquad \text{($N$-subjettiness, fail),}
\end{alignat}
and later combine them into one ROC curve. This precise condition is
represented by the more generic Eq.\eqref{eq:vars_nsub}.  In
Fig.~\ref{fig:performance} we show the corresponding ROC curves for a
successively improved tagger.\bigskip

Finally, we can replace the deterministic clustering history from the
usual jet algorithm with a set of \textsc{Qjets} histories with large
global weights $\Omega^{(\alpha)}$ defined in Eq.\eqref{eq:qjets2}
for $\alpha = 0.1$. This way we avoid cases where the deterministic clustering
history entering the top tagging algorithm  is misled during the
independent evaluation of splittings in the usual jet algorithm. When
defining jets as analysis objects for a hard process this does not
pose a problem, but for subjet analyses it can have an effect.

%---------------------------------------------------------------------
\begin{table}[b!]
\begin{tabular}{l|r|r}
 & $t\bar{t}$ & QCD \\ \hline
default HTT & 0.337 & 0.0212 \\ \hline
$\eps_\text{Qjets} > 0.1$ & 0.435 & 0.0318 \\ 
$\eps_\text{Qjets} > 0.2$ & 0.384 & 0.0231 \\ 
$\eps_\text{Qjets} > 0.3$ & 0.341 & 0.0174 \\ 
$\eps_\text{Qjets} > 0.4$ & 0.298 & 0.0123 \\
$\eps_\text{Qjets} > 0.5$ & 0.250 & 0.0089 \\
$\eps_\text{Qjets} > 0.6$ & 0.212 & 0.0064 \\
$\eps_\text{Qjets} > 0.7$ & 0.163 & 0.0036 \\
$\eps_\text{Qjets} > 0.8$ & 0.118 & 0.0021 \\
$\eps_\text{Qjets} > 0.9$ & 0.064 & 0.0007 \\ \hline
\end{tabular} \hspace*{0.1\textwidth}
\begin{tabular}{l|r|r}
 & $t\bar{t}$ & QCD \\ \hline
default HTT & 0.465 & 0.0489 \\ \hline
$\eps_\text{Qjets} > 0.1$ & 0.524 & 0.0661 \\
$\eps_\text{Qjets} > 0.2$ & 0.447 & 0.0461 \\
$\eps_\text{Qjets} > 0.3$ & 0.388 & 0.0342 \\
$\eps_\text{Qjets} > 0.4$ & 0.336 & 0.0245 \\
$\eps_\text{Qjets} > 0.5$ & 0.281 & 0.0168 \\
$\eps_\text{Qjets} > 0.6$ & 0.236 & 0.0118 \\ 
$\eps_\text{Qjets} > 0.7$ & 0.181 & 0.0062 \\ 
$\eps_\text{Qjets} > 0.8$ & 0.133 & 0.0032 \\ 
$\eps_\text{Qjets} > 0.9$ & 0.069 & 0.0009 \\ \hline
\end{tabular}

\caption{Tagging efficiencies for $p_T > 200$~GeV (left) and $p_T >
  600$~GeV (right). $\eps_\text{Qjets}$ is defined as the number
  of \textsc{Qjets} tags per number of \textsc{Qjets} runs. For this
  table we test 10.000 fat jets with 100 \textsc{Qjets} iterations.}
\label{tab:qjets}
\end{table}
%---------------------------------------------------------------------

Our analysis is based on 100 \textsc{Qjets} histories per fat jet.  In
Tab.~\ref{tab:qjets} we show their signal and background efficiency if
required to lead to individual top tags. As the reference value we use
the default \textsc{HEPTopTagger} with fixed mass windows. Based on
100 \textsc{Qjets} histories we then define the fraction
$\eps_\text{Qjets}$ of histories which lead to a top tag with the
default tagging setup. We see that for moderately boosted tops the
deterministic signal tagging efficiency can be reproduced by requiring
$30\%$ of the \textsc{Qjets} histories to deliver a positive tag. The
corresponding mis-tag probability is slightly reduced compared to the
deterministic tagger. For harder tops the corresponding value is
around $\eps_\text{Qjets} > 20\%$, with no improvement in the
background rejection.\bigskip

%---------------------------------------------------------------------
\begin{figure}[t!]
  \includegraphics[width=0.4\textwidth]{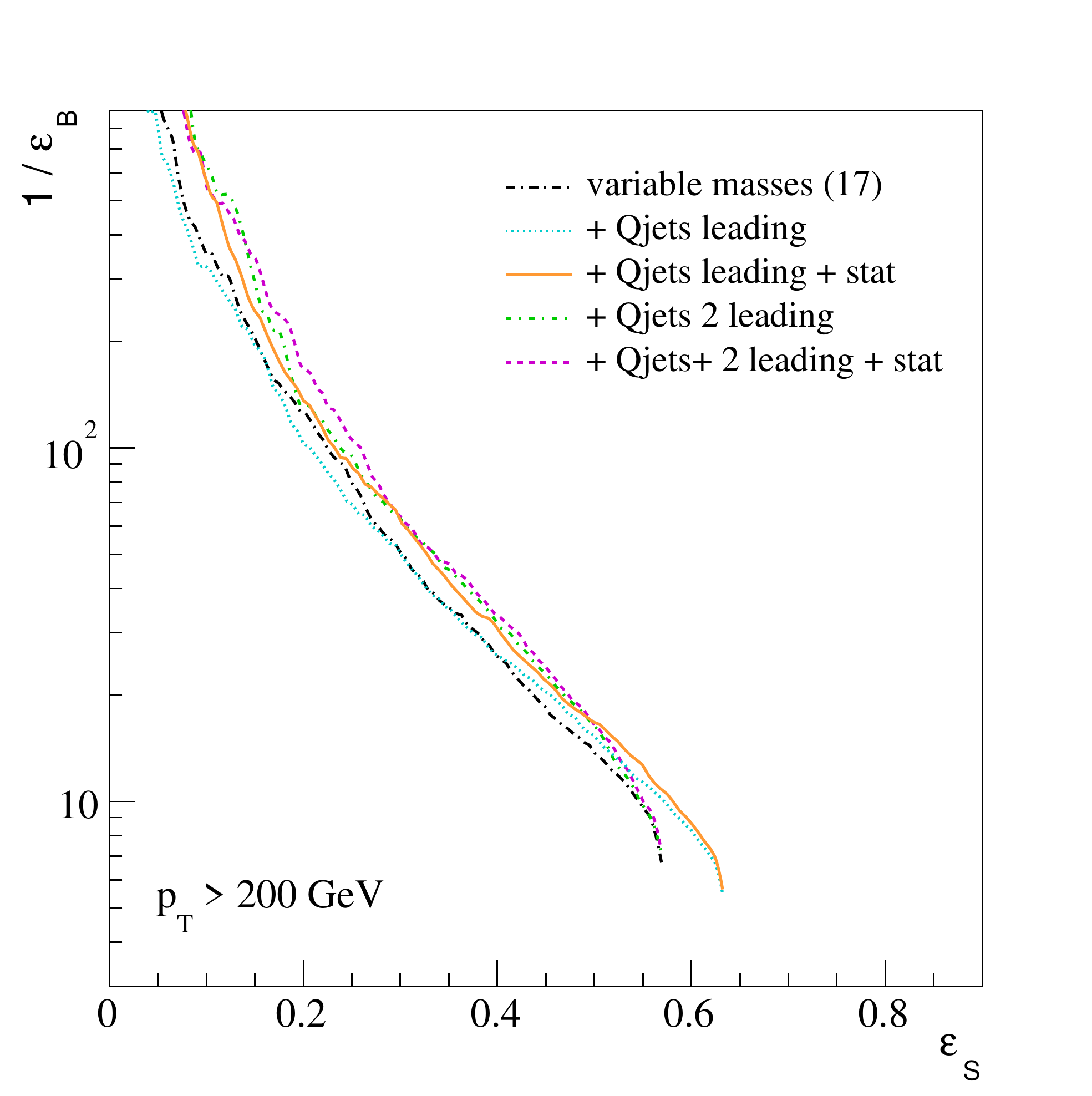}
  \hspace*{0.1\textwidth}
  \includegraphics[width=0.4\textwidth]{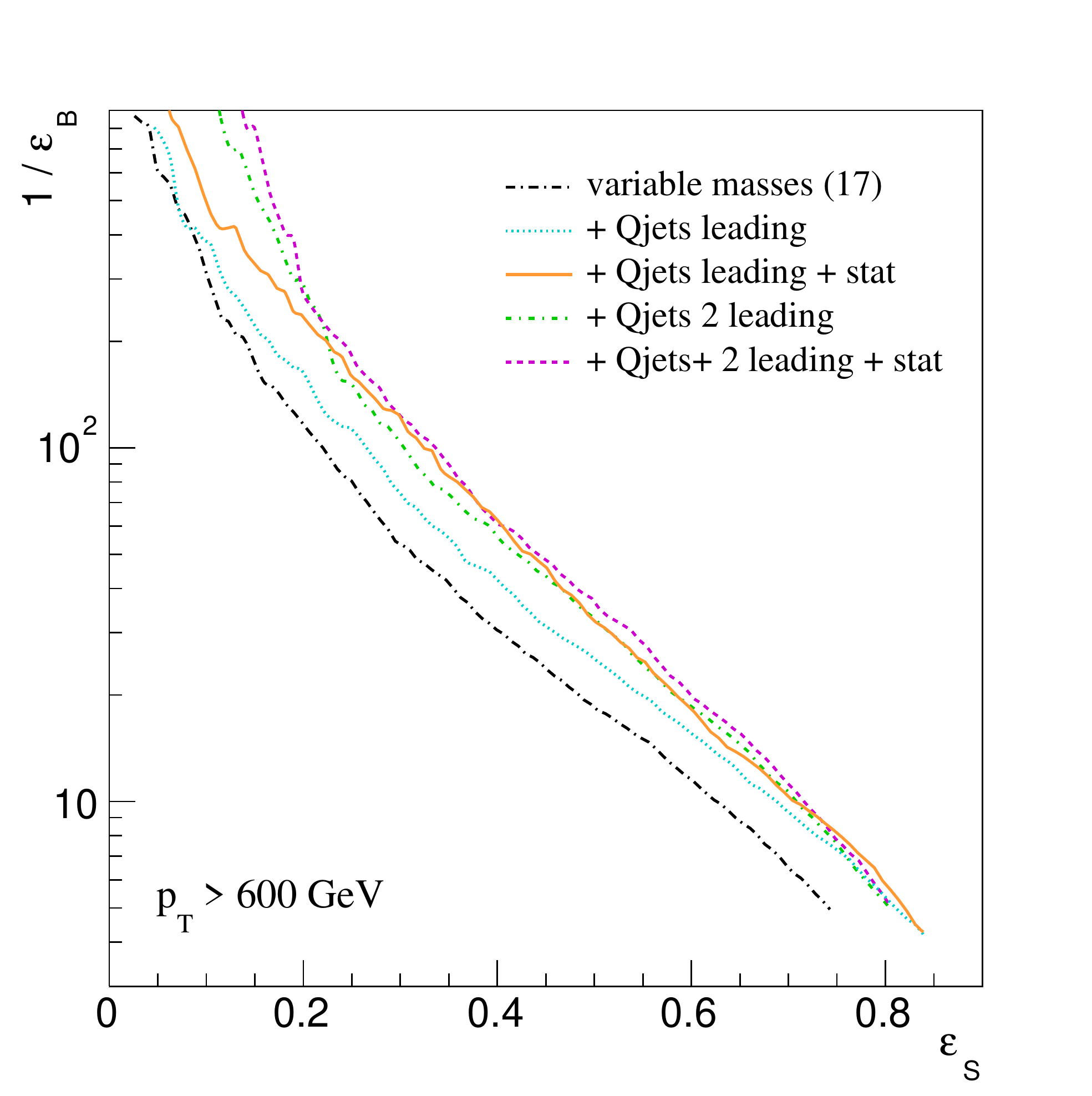}
\caption{Performance of the \textsc{HEPTopTagger2} for $t\bar{t}$
  production in the Standard Model. For $p_{T,t} > 200~\gev$ and
  $p_{T,t} > 600~\gev$ we we focus on different \textsc{Qjets} setups,
  based on a more basic multivariate tagger without optimalR and
  $N$-subjettiness.}
  \label{fig:performance_qjets}
\end{figure}
%---------------------------------------------------------------------

As discussed in Sec.~\ref{sec:optr} \textsc{Qjets} offers two
strategies to improve the top tagger. To maximize the improvement in
the tagging performance and to limit the CPU time we base the
multivariate analysis on the tagged history with the largest global weight.  As
additional parameters we include the value of $\eps_\text{Qjets}$ as
well as the mean and variance of the $m_\text{rec}$ distribution with the
100 \textsc{Qjets} entries, symbolically denoted as
$\{ m_\text{rec}^\text{Qjets} \}$. For the BDT analysis the variables are
\begin{alignat}{2}
&\{ \ m_\text{rec},  f_\text{rec},
R_\text{opt} - R_\text{opt}^{(\text{calc})}, m_\text{fat},  
\tau_N, \tau_N^\text{(filt)},
\eps_\text{Qjets}, \{ m_\text{rec}^\text{Qjets} \}  \; \} 
\qquad \text{(\textsc{Qjets})}
\end{alignat}
As usual, all variables from the tagger are evaluated for the
optimized $R$ size and the clustering history with the largest global
weight. The additional improvement is shown in
Fig.~\ref{fig:performance}.\bigskip

Because \textsc{Qjets} offers a variety of improvements to the tagger,
we study different setups based on the stage with multivariate mass
windows in Fig.~\ref{fig:performance_qjets}. We start by replacing the
deterministic C/A output with the most likely \textsc{Qjets} history
and including $\eps_\text{Qjets}$ in the multivariate analysis. This
leads to a moderate improvement of the tagger at large transverse
momenta and at large signal efficiencies. Adding the statistical
information from the $\eps_\text{Qjets} \times 100$ entries in the
$m_\text{rec}$ information leads to a sizeable improvement over a wide
range of signal efficiencies. This is the mode we use for the $Z'$
analysis as well as in Fig.~\ref{fig:performance}.

Next, we add the second-best \textsc{Qjets} history to the tagger,
such that the multivariate tagger (including $\eps_\text{Qjets}$) is
free to construct a criterion based on one or two tags in the two best
\textsc{Qjets} histories. For most of the ROC curves this comparably
simple approach is as successful as the full statistical
information. Finally, adding the statistical information on the
$m_\text{rec}$ distribution leads to a mild improvement.

%%%%%%%%%%%%%%%%%%%%%%%%%%%%%%%%%%%%%%%%%%%%%%%%%%%%%%%%%%%%%%%%%%%%%%
\subsection*{Interface}

To apply the \textsc{HEPTopTagger} algorithm to a fat C/A jet
constructed with \textsc{FastJet3}~\cite{fastjet}, the only necessary
steps are executing the default constructor
\texttt{HEPTopTagger(fastjet::PseudoJet jet)} followed by running the
tagger using \texttt{void run()}. This will analyze the fat
jet using the optimalR procedure with the default settings given in
Tab.~\ref{tab:htt_params}.  The available operation modes are shown in
Tab.~\ref{tab:htt_modes}. All configurable parameters are listed in
Tab.~\ref{tab:htt_params}. Functions to retrieve results are presented
in Tab.~\ref{tab:htt_res}.

\texttt{QHTT()} sets up the \textsc{Qjets} mode. It is applied to a
fully configured \textsc{HEPTopTagger} by \texttt{void
  run(HEPTopTagger htt)}. All configurable parameters are given in
Tab.~\ref{tab:qjets_params}. A list of functions to access the results
is presented in Tab.~\ref{tab:qjets_res}.

In addition, we provide a framework for the calculation of
Fox--Wolfram~moments that relies on an existing installation of
\textsc{GSL}~\cite{gsl}. While the constructor
\texttt{FWM(vector<fastjet::PseudoJet> jets)} allows the calculation
of Fox--Wolfram~moments for a given set of jets,
\texttt{FWM(HEPTopTagger htt, unsigned selection)} uses the $b$,
$W_1$, and $W_2$ momenta from the \textsc{HEPTopTagger} run and
calculates the Fox--Wolfram~moments in the top rest frame. The boost
axis $\vec{a}$ itself can be included~\cite{heptop4}. Subsets of these
four vectors can be set via \texttt{unsigned selection}, as a sequence
of 0 or 1 in the order $abW_1W_2$. In Tab.~\ref{tab:fwm_res} we show
how to extract the Fox--Wolfram~moment of a given order of the Legendre polynomials.

Finally, we include an example class \texttt{LowPt()} for a fixed
low-$p_T$ mode working point returning a tagging decision including
the set low-$p_T$ mode by \texttt{is\_tagged(HEPTopTagger)}.

%---------------------------------------------------------------------
\begin{table}[t!]
\begin{small}
\begin{tabular}{p{6 cm} | p{10 cm}}
\hline
name & description \\
\hline
\texttt{EARLY\_MASSRATIO\_SORT\_MASS} & apply the 2D mass plane requirements, then select the candidate which minimizes $|m_\text{cand}-m_t|$ \\
\texttt{LATE\_MASSRATIO\_SORT\_MASS} & select the candidate which minimizes $|m_\text{cand}-m_t|$ \\
\texttt{EARLY\_MASSRATIO\_SORT\_MODDJADE} & apply the 2D mass plane requirements, then select the candidate with the highest modified Jade distance \\
\texttt{LATE\_MASSRATIO\_SORT\_MODDJADE} & select the candidate with the highest modified Jade distance \\
\texttt{TWO\_STEP\_FILTER} & only analyze the candidate built with the highest $p_{T,t}$ after unclustering \\
\hline
\end{tabular}
\end{small}
\caption{\textsc{HEPTopTagger} working modes.}
\label{tab:htt_modes}
\end{table}
%---------------------------------------------------------------------

%---------------------------------------------------------------------
\begin{table}[b!]
\begin{small}
  \begin{tabular}{p{6.3cm} | p{4.1cm} | p{5.5cm}}
  \hline
  name & default & description \\
  \hline

  general: & & \\
  \texttt{do\_optimalR(bool)} & \texttt{true} & use optimalR approach \\[2mm] 

  unclustering: & & \\
  \texttt{set\_mass\_drop\_threshold(double)} & 0.8 & mass drop threshold \\
  \texttt{set\_max\_subjet\_mass(double)} & 30 & max subjet mass for unclustering \\[2mm]
  
  filtering: & & \\
  \texttt{set\_filtering\_R(double)} & 0.3 & max subjet distance for filtering \\
  \texttt{set\_filtering\_n(unsigned)} & 5 & max subjet number for filtering \\
    \texttt{set\_filtering\_minpt\_subjet(double)} & 0. & min subjet $p_T$ for filtering \\
    \texttt{set\_filtering\_jetalgorithm(\newline \hspace*{6ex} fastjet::JetAlgorithm)} & \texttt{cambridge\_algorithm} & jet algorithm for filtering\newline \\[2mm]

  reclustering: & & \\
  \texttt{set\_reclustering\_jetalgorithm(\newline \hspace*{6ex} fastjet::JetAlgorithm)}& \texttt{cambridge\_algorithm} & jet algorithm for reclustering\newline \\[2mm]

  candidate selection: & & \\
  \texttt{set\_mode(enum)} & \texttt{EARLY\_MASSRATIO\_SORT\_MASS} & run mode, see Tab.~\ref{tab:htt_modes} \\
  \texttt{set\_mt(double)} & 172.3 & true top mass\\
  \texttt{set\_mw(double)} & 80.4 & true W mass\\
  \texttt{set\_top\_mass\_range(double, double)} & 150, 200 & top mass window \\    
 \texttt{set\_fw(double)} & 0.15 & width of A--shaped bands $f_W$\\
  \texttt{set\_mass\_ratio\_range( \newline \hspace*{6ex} double,
  double)} & $(1-f_W)\, m_W/m_t=0.397$ \newline $(1+f_W)\, m_W/m_t=0.537$ & width of cut in 2D mass plane \\
  \texttt{set\_mass\_ratio\_cut(double, \newline \hspace*{6ex}
  double, double)} & 0.35, 0.2, 1.3 & boundaries in 2D mass plane\\
  \texttt{set\_top\_minpt(double)} & 200 & min $p_{T,t}$ consistency cut \\
  [2mm]

  pruning: & & \\
  \texttt{set\_pruning\_zcut(double)} & 0.1 & $z_\text{cut}$ for pruned mass $m_\text{prune}$ \\ 
  \texttt{set\_pruning\_rcut\_factor(double)} & 0.5 & $r_\text{cut}$ for pruned mass $m_\text{prune}$ \\[2mm] 
  
  optimalR: & & \\
  \texttt{set\_optimalR\_max(double)} & size of the input fat jet & max jet size \\
  \texttt{set\_optimalR\_min(double)} & 0.5 & min jet size \\
  \texttt{set\_optimalR\_step(double)} & 0.1 & step size (multiple of 0.1) \\
  \texttt{set\_optimalR\_threshold(double)} & 0.2 & optimalR mass
                                                    threshold \\[2mm]

  calculation of $R_\text{opt}^\text{(calc)}$: & & \\
  \texttt{set\_filtering\_optimalR\_calc\_R(double)} & 0.2 & max subjet
                                                        distance for filtering \\
  \texttt{set\_filtering\_optimalR\_calc\_n(unsigned)} & 10 & max subjet
                                                        number for filtering \\
  \texttt{set\_optimalR\_calc\_fun(double (*f)(double))} &
                                                           $327/p_{T,\text{filt}}$
                 & dependency of $R_\text{opt}^\text{(calc)}$ on $p_{T,\text{filt}}$ \newline \\[2mm]

  optimalR type: & & \\
  \texttt{set\_optimalR\_type\_top\_mass\_range(double, double)} & 150. 200. & mass
                                                                range
                                                                for
                                                                optimalR type 1  \\
  \texttt{set\_optimalR\_type\_f\_rec(double)}& 0.175 & max $f_\text{rec}$ for optimalR type 1\\
  \texttt{set\_optimalR\_type\_max\_diff(double)} & 0.3 &
                                                         max $R_\text{opt}
                                                         -
                                                           R_\text{opt}^\text{(calc)}$ for
                                                                optimalR type 1\\[2mm]

  $N$-subjettiness: & & \\
  \texttt{set\_filtering\_optimalR\_pass\_R(double)} & 0.2 &
                                                             $R_\text{filt}$ for optimalR type 1\\
  \texttt{set\_filtering\_optimalR\_pass\_n(unsigned)} & 5 &  $N_\text{filt}$ optimalR type 1\\
  \texttt{set\_filtering\_optimalR\_fail\_R(double)} & 0.3 &
                                                             $R_\text{filt}$
                                                             for optimalR type 0\\
  \texttt{set\_filtering\_optimalR\_fail\_n(unsigned)} & 3 &
                                                             $N_\text{filt}$ for
                                                             optimalR type 0\\
  
\hline
\end{tabular}
\end{small}
\caption{Additional parameters of the \textsc{HEPTopTagger} algorithm. All functions have a return type of \texttt{void}.}
\label{tab:htt_params}
\end{table}
%---------------------------------------------------------------------

%---------------------------------------------------------------------
\begin{table}[t]
\begin{small}

\begin{tabular}{p{7.7cm} | p{8cm}}
  \hline
  name & description \\
  \hline

  \texttt{bool is\_maybe\_top()} & top mass window requirement passed? \\
  \texttt{bool is\_masscut\_passed()} & 2D mass plane requirements passed? \\
  \texttt{bool is\_minptcut\_passed()} & candidate $p_{T,t}$ threshold passed? \\
  \texttt{bool is\_tagged()} & top mass window, 2D mass plane
                               requirement, and $p_{T,t}$ threshold passed? \\
  \texttt{double delta\_top()} & $|m_\text{rec}-m_t|$ \\
  \texttt{double djsum()} & modified Jade distance \\
  \texttt{double pruned\_mass()} & pruned top mass \\
  \texttt{double unfiltered\_mass()} & mass of the triplet of subjets after unclustering before filtering \\
  \texttt{double f\_rec()} & minimal $|(m_{ij}/m_\text{rec})/(m_W/m_t) - 1|$ \\
  \texttt{const PseudoJet \& t()} & top candidate 4-vector \\
  \texttt{const PseudoJet \& b()} & subjet corresponding to the $b$ \\
  \texttt{const PseudoJet \& W()} & combined subjets corresponding to the $W$ \\
  \texttt{const PseudoJet \& W1()} & leading subjet from the $W$ \\
  \texttt{const PseudoJet \& W2()} & sub-leading subjet from the $W$ \\
  \texttt{const std::vector<PseudoJet> \& top\_subjets()} & three subjets from the top, ordered: $b$, $W_1$, $W_2$ \\ 
  \texttt{const PseudoJet \& j1()} & leading subjet \\
  \texttt{const PseudoJet \& j2()} & sub-leading subjet \\
  \texttt{const PseudoJet \& j3()} & sub-sub-leading subjet \\
  \texttt{const std::vector<PseudoJet> \& top\_hadrons()} & all top constituents \\
  \texttt{const std::vector<PseudoJet> \& hardparts()} & hard subtructures after unclustering, sorted by $p_T$ \\
  \texttt{const PseudoJet \& fat\_inital()} & original fat jet (after
                                              \textsc{Qjets} reclustering) \\  
  \texttt{const PseudoJet \& fat\_Ropt()} & fat jet reduced to $R_\text{opt}$ \\
  \texttt{void get\_setting()} & print settings to \texttt{stdout} \\ 
  \texttt{void get\_info()} & print tagger information to \texttt{stdout} \\
  \texttt{\textsc{HEPTopTagger} HTTagger(unsigned i)} & \textsc{HEPTopTagger} candidate for a distance parameter $R = i/10$. By default all functions above return values at $R=R_\text{opt}$. This function accesses candidates for different values of $R$. \\
  \texttt{double Ropt()} & $R_\text{opt}$ \\
  \texttt{double Ropt\_calc()} & $R_\text{opt}^\text{(calc)}$ \\
  \texttt{int optimalR\_type()} & result of set optimalR working
                                  point. 1 = pass, 0 = fail \\

  \texttt{double nsub\_unfiltered(int order,
  fastjet::contrib::Njettiness::AxesMode axes =
  fastjet::contrib::Njettiness::kt\_axes, double beta = 1., double R0 =
  1.);} & $N$--subjettiness $\tau_i$ for the unfiltered fat jet \\
  \texttt{double nsub\_filtered(int order,
  fastjet::contrib::Njettiness::AxesMode axes =
  fastjet::contrib::Njettiness::kt\_axes, double beta = 1., double R0 =
  1.);} & $N$--subjettiness $\tau^\text{(filt)}_i$ for the fat jet after
  filtering depending on \texttt{optimalR\_type()}. \\
  \texttt{double q\_weight()} & weight of used \textsc{Qjets} history\\

  \hline
\end{tabular}
\end{small}
\caption{Functions to retrieve results of the \textsc{HEPTopTagger} algorithm.}
\label{tab:htt_res}
\end{table}
%---------------------------------------------------------------------

%---------------------------------------------------------------------
\begin{table}[b!]
\begin{small}
\begin{tabular}{p{6.3cm} | p{2.6cm} | p{6.8cm}}
  \hline
  name & default & description \\
  \hline

 \texttt{set\_iterations(unsigned)} & 100 & number of \textsc{Qjets}
                                            iterations\\
\texttt{set\_q\_zcut(double)} & 0.1 & $z_\text{cut}$ for pruning in
                                      \textsc{Qjets}\\
\texttt{set\_q\_dcut\_fctr(double)} & 0.5 & $D_\text{cut}$ factor for
                                            pruning in
                                           \textsc{Qjets}\\
\texttt{set\_q\_exp(double a, double b)} & 0., 0. (C/A) & set distance
                                                     measure for
                                                     \textsc{Qjets} \newline
                                                     $d_{ij} =
                                                     \min(p_{T,i},p_{T,j})^a
                                                     \,
                                                     \max(p_{T,i},p_{T,j})^b
                                                     \, R_{ij}^2$\\
\texttt{set\_q\_rigidity(double)} & 0.1 & rigidity $\alpha$ for
                                           \textsc{Qjets}\\
\texttt{set\_q\_truncation\_fctr(double)} & 0. & threshold for
                                                  merging probability $\omega_{ij}$
                                                  in 
                                           \textsc{Qjets}\\
  
\hline
\end{tabular}
\end{small}
\caption{Parameters of the \textsc{Qjets} frame for the \textsc{HEPTopTagger}. All functions have a return type of \texttt{void}.}
\label{tab:qjets_params}
\end{table}
%---------------------------------------------------------------------

%---------------------------------------------------------------------
\begin{table}[t]
\begin{small}

\begin{tabular}{p{5.7cm} | p{10cm}}
\hline
name & description \\
\hline
\texttt{HEPTopTagger leading()} & HEPTopTagger with leading tagged
 history\\
\texttt{HEPTopTagger subleading()} & HEPTopTagger with subleading tagged
 history\\
\texttt{double weight\_leading()} & \textsc{Qjets} weight of the
 leading tagged history\\
\texttt{double weight\_subleading()} & \textsc{Qjets} weight of
 the subleading tagged history\\
\texttt{double eps\_q()} & $\eps_\text{Qjets}$\\
\texttt{double m\_mean()} & $\langle m \rangle$ for the tagged histories\\
\texttt{double m2\_mean()} & $\langle m^2 \rangle$ for the tagged histories\\

\hline
\end{tabular}
\end{small}
\caption{Functions to retrieve results of the \textsc{Qjets} frame.}
\label{tab:qjets_res}
\end{table}
%---------------------------------------------------------------------

%---------------------------------------------------------------------
\begin{table}[hbt]
\begin{small}\begin{tabular}{p{5.7cm} | p{10cm}}
\hline
name & description \\
\hline
 \texttt{double U(unsigned)} & FWM of given order with unit weight \\
 \texttt{double Pt(unsigned, fastjet::PseudoJet=$(0., 0., 1., 0.)$)}
 & FWM of given order with $p_T$ weight relative to the given
 reference vector. \\
 
\hline
\end{tabular}
\end{small}
\caption{Functions to retrieve Fox--Wolfram moments.}
\label{tab:fwm_res}
\end{table}
%---------------------------------------------------------------------

\clearpage

%%%%%%%%%%%%%%%%%%%%%%%%%%%%%%%%%%%%%%%%%%%%%%%%%%%%%%%%%%%%%%%%%%%%%%

\end{document}